\documentclass[10pt,letterpaper]{article}
\usepackage[top=0.85in,left=2.75in,footskip=0.75in]{geometry}

\usepackage{amsmath,amssymb}

\usepackage{changepage}

\usepackage{textcomp,marvosym}

\usepackage{cite}

\usepackage{nameref}

\usepackage[right]{lineno}

\usepackage[nopatch=eqnum]{microtype}
\DisableLigatures[f]{encoding = *, family = * }

\usepackage{array}

\usepackage{siunitx}

\usepackage[svgnames, table]{xcolor} 

\usepackage{nicematrix} 

\usepackage{tabularx}

\usepackage[colorlinks=true, linkcolor=blue, citecolor=blue, filecolor=blue,
runcolor=blue, urlcolor = black]{hyperref}

\newcommand{\atznum}[2][2]{
\num[round-precision=#1,round-mode=figures,
     scientific-notation=true]{#2}
}

\newcommand{\dvol}{\delta{V}} 

\newcommand{\atzadjustf}[1]{\substack{#1\\\mbox{ }}}

\newcolumntype{+}{!{\vrule width 2pt}}

\newlength\savedwidth

\raggedright
\usepackage[aboveskip=1pt,labelfont=bf,labelsep=period,
justification=raggedright,singlelinecheck=off]{caption}

\makeatletter
\renewcommand{\@biblabel}[1]{\quad#1.}
\makeatother

\usepackage{lastpage,fancyhdr,graphicx}
\usepackage{epstopdf}
\fancyheadoffset[L]{2.25in}
\fancyfootoffset[L]{2.25in}

\definecolor{issuePJA_color}{rgb}{1.0,0.0,0.0}

\definecolor{commentPJA_color}{rgb}{1.0,0.0,0.8}

\definecolor{revB_color}{rgb}{0.6,0.4,0.0}
\newcommand{\revB}[1]{#1}

\definecolor{rev_color}{rgb}{0.6,0.0,0.0}
\newcommand{\rev}[1]{#1}

\definecolor{atz_table1}{rgb}{0.85, 0.85, 0.85}
\definecolor{atz_table2}{rgb}{0.8, 0.8, 0.8}
\definecolor{atz_table3}{rgb}{0.75, 0.75, 0.75}

\newcommand*{\mb}[1]{\mathbf{#1}} 
\newcommand*{\bsy}[1]{\boldsymbol{#1}}

\newcommand{\gthm}{\mb{g}_{thm}} 

\definecolor{atzmlabel_color}{rgb}{0.0,0.0,0.0}
\newcommand{\atzmfont}{\fontsize{8}{9}} 
\newcommand{\atzmlabel}[1]{\mbox{\atzmfont \textcolor{atzmlabel_color}{#1}}}

\begin{document}
\vspace*{0.2in}

\begin{flushleft}
{\Large \textbf\newline{\rev{Protein Drift-Diffusion in Membranes
with Non-equilibrium Fluctuations arising from Gradients in
Concentration or Temperature. }}}

D. Jasuja\textsuperscript{1},
P. J. Atzberger\textsuperscript{1*},
\\
\bigskip
\textbf{1} 
Department of Mathematics, Department of Mechanical Engineering, 
University of California Santa Barbara (UCSB), Santa Barbara, CA, USA
\bigskip

* atzberg@gmail.com; \url{https://atzberger.org/}

\end{flushleft}
\section*{Abstract}
\rev{We investigate proteins within heterogeneous cell membranes where
non-equilibrium phenomena arises from spatial variations in concentration and
temperature.  We develop simulation methods building on non-equilibrium
statistical mechanics to obtain stochastic hybrid continuum-discrete
descriptions which track individual protein dynamics, spatially varying
concentration fluctuations, and thermal exchanges.  We investigate biological
mechanisms for protein positioning and patterning within membranes and factors
in thermal gradient sensing.  We also study the kinetics of Brownian motion of
particles with temperature variations within energy landscapes arising from
heterogeneous microstructures within membranes.  The introduced approaches
provide self-consistent models for studying biophysical mechanisms involving
the drift-diffusion dynamics of individual proteins and energy exchanges and
fluctuations between the thermal and mechanical parts of the system.  The
methods also can be used for studying related non-equilibrium effects in other
biological systems and soft materials.
}

\section*{Author Summary}
We introduce theoretical frameworks and computational simulation methods for
modeling and investigating protein dynamics within heterogeneous membranes in
non-equilibrium regimes.  A hybrid discrete-continuum approach is developed
allowing for tracking individual proteins and their coupling to
spatial variations in concentration and temperature captured by continuum
fluctuating fields.  Investigations are performed of biological processes and
related phenomena arising from fluctuations, gradients in concentration,
variations in temperature, and other non-equilibrium effects.

\section*{Introduction} 

\rev{Cellular membranes are heterogeneous mixtures of
proteins, lipids, and other small molecules~\cite{Alberts2002a,
singer1972fluid,goni2014basic,ali2023review,krause2014structural,nicolson2023fluid}.
The spatial-temporal organization of membrane proteins and their kinetics 
play important roles in biological 
functions~\cite{ramos2022biomembrane,sunshine2017membrane,
roy2023lipid,owen2009quantitative}. This often involves 
non-equilibrium processes~\cite{bustamante2005nonequilibrium,
fang2019nonequilibrium,fang2020nonequilibrium} and related phenomena breaking 
detailed-balance~\cite{gnesotto2018broken,battle2016broken}. 
This includes active transport by motor proteins~\cite{atzberger2006brownian,
peskin1995coordinated,sweeney2018motor}, mitochondrial ATP
synthesis~\cite{rajagopal2019transient,lane2018hot,fahimi2022hot}, ion
exchanges by pumps and facilitated diffusion~\cite{carruthers1991mechanisms,Alberts2002a}, 
active cytoskeletal forces~\cite{mogilner1996cell,fletcher2010cell}, and 
other mechanisms~\cite{Alberts2002a,bustamante2005nonequilibrium}.
A fundamental challenge in studying many of these cellular processes, and
related \textit{in vitro} experimental systems, is to gain insights into the membrane 
proteins and their spatial-temporal distribution and kinetics.
This is impacted by drift-diffusion 
dynamics and reactions coupled to signals originating both 
within the cell and from the surrounding 
environment~\cite{mashanov2021heterogeneity,Alberts2002a,
chai2022heterogeneous,jeon2016protein}.
Cellular and environmental factors can include local variations and gradients 
in concentrations of signaling molecules~\cite{he2016dynamic,
sakamoto2023heterogeneous,
bressloff2019protein}
and variations in external temperatures~\cite{chen2021heat,rings2011theory}.  
Recent \textit{in vitro} experiments also have been developed to probe 
the interactions of proteins and synthetic particles with 
local concentration and temperature gradients, and to manipulate them at the 
single-molecule level ~\cite{piazza2008thermophoresis,talbot2017thermal,
chen2021heat,garner2013cell,Jiang2010,Garcia2005}.}

\rev{Capturing non-equilibrium effects important in these biological systems and
experimental assays poses several challenges for theoretical modeling and
practical simulation.  This includes the need to capture the drift-diffusion
dynamics of proteins as they move through regions in the membrane having
different concentrations, temperatures, or other types of heterogeneity.  In
addition to the dynamics of the protein motions, the concentration and
temperature fields can change over time from exchanges of mass and energy.  For
continuum mechanics descriptions at such spatial-temporal scales, both local
concentrations and temperatures can spontaneously fluctuate from under-resolved
smaller-scale exchanges.}

\revB{Most theoretical modeling and simulation methods treat homogeneous systems.
This includes Brownian-Stokesian Dynamics~\cite{Banchio2003,McCammon1988},
Coarse-grained approaches with Langevin thermostats~\cite{Doi2013,
Deserno2009multiscale,noid2013perspective}, and continuum mechanics
formulations such as Stochastic Immersed Boundary Methods
(SIBMs)~\cite{Atzberger2007,AtzbergerSoftMatter2016},
related fluctuating hydrodynamic approaches~\cite{martinez2024finite,de2016finite,AtzbergerSurfFluctHydro2022},
and Stochastic Eulerian Lagrangian Methods
(SELMs)~\cite{AtzbergerSELM2011,AtzbergerTabak2015}.  These computational
simulation methods are based on continuum hydrodynamic descriptions and
statistical mechanics primarily in regimes at thermodynamic equilibrium.}

\rev{Related work treating non-equilibrium regimes include recent theoretical and
simulation studies of Hot Brownian Motion and Soret effects that treat
diffusion of particles within temperature
gradients~\cite{kohler2016soret,rings2010hot}.  This has been modeled by using
temperature dependent viscosities~\cite{rings2010hot}, renormalized
diffusivities~\cite{rings2011theory}, or through molecular dynamics simulations
of a particle in Lennard-Jones fluids
\cite{schachoff2015hot,chakraborty2011generalised,chakraborty2018orientational}.
\revB{There also has been some work using fluctuating hydrodynamics to derive
generalized Langevin equations for Hot Brownian Motions for translational and
rotational motions~\cite{rings2012rotational,chakraborty2011generalised}
and for proteins that can react to or generate curvature in 
membranes~\cite{tozzi2019out,sigurdsson2013hybrid}.}
Most of these approaches assume regimes with time-scale separation where they
can reduce the descriptions to effective tensors for a single particle and do
not track environmental changes in the temperature or concentration fields and their
fluctuations. }
 
\rev{We develop here non-equilibrium statistical mechanics approaches that use
hybrid discrete-continuum descriptions.  We capture both the individual protein
drift-diffusion dynamics and the spatial-temporal evolution and fluctuations
arising from spatially varying concentration and temperature fields within the
membrane.  We couple proteins tracked at the single-molecule level with these
fields.  We also develop stochastic numerical methods to obtain practical simulation
approaches.  We consider in this initial work the case when the membranes 
are treated as static without shape undulations where the geometry remains
approximately flat.  We develop methods for 
capturing non-equilibrium effects impacting the drift-diffusion 
dynamics of individual proteins within heterogeneous
environments.}

\rev{We demonstrate how our approaches can be used to 
investigate phenomena in membrane-protein systems that include
(i) how concentration gradients and kinetics 
drive the spatial organization of proteins, 
(ii) the roles of fluctuations in the encoding
of signals by proteins to sense  
external thermal gradients, and (iii) how localized 
laser heating, as in \textit{in vitro} experiments,
can probe and impact protein escape kinetics from local 
energy wells within heterogeneous membranes.  
Our introduced simulation methods also can be used
to investigate other non-equilibrium
phenomena for systems where 
significant roles are played by
particle drift-diffusion dynamics coupled 
to local variations and fluctuations 
in concentration or temperature.}

\section*{Materials and methods}

\subsection*{Membrane-Protein System: Stochastic Non-Equilibrium Model}

We model the membrane and proteins using a hybrid stochastic continuum-discrete
approach.  For the drift diffusion dynamics of an individual protein, we use
\begin{eqnarray}
\label{equ_full_model}
\frac{d\mb{X}}{dt} & = & \mb{M}_{\tiny XX} \mb{F}_X + k_B \theta_P\left(\nabla_X
\cdot \mb{M}_{\tiny XX}\right) + \mb{H}_{\tiny thm,X}. \;\hspace{0.4cm}
\end{eqnarray}
\revB{The protein location within the membrane is denoted by $\mb{X}$, the
forces by $\mb{F}_X$, the temperature by $\theta_P$, and thermal fluctuations
by $\mb{H}_{\tiny thm,X}$.  The over-damped kinematic hydrodynamic response to
an applied force is given by the mobility $\mb{M}_{\tiny
XX}$~\cite{atzberger2007note,McCammon1988,Doi2013}.  The $k_B$ denotes the
Boltzmann constant~\cite{Reichl1997}. }

\revB{For convenience, we also alternatively will refer to the combined
contributions of the fluctuations to the protein using notation $\mb{G}_{\tiny
thm,X} = k_B \theta_P\left(\nabla_X \cdot \mb{M}_{\tiny XX}\right) +
\mb{H}_{\tiny thm,X}$.  The divergence term in equation~\ref{equ_full_model}
arises from the diffusivity depending on the mobility $\mb{M}_{\tiny XX}$ which
has a spatial dependence~\cite{atzberger2007note,Reichl1997}.  In the case of
conservative forces, we have $\mb{F}_X = -\partial_{X} U$.  We discuss how the
thermal fluctuations are determined and other details below. }

\revB{For tracking the concentration of a chemical species within the membrane, 
we use the continuum field $c(x,t) = c_0q(x,t)$.  The $q$ has the dynamics}
\begin{eqnarray}
\label{equ_conc}
\frac{\partial q(x,t)}{\partial t} &=& 
\mbox{\small div}\left(
\bar{\kappa} \nabla q(x,t)
\right)
-\mbox{\small div}\left({-\frac{1}{\gamma} q(x,t) \nabla \Phi(x;\mb{X}) }\right)
+ \mb{g}_{\tiny thm,q}.
\hspace{0.4cm} 
\end{eqnarray}
\revB{The total concentration is given by $c_0$. 
The diffusivity is denoted by $\bar{\kappa}$ and satisfies the
Stokes-Einstein relation
$\bar{\kappa} = \theta_C/\gamma$~\cite{Reichl1997}.
The $\theta_C$ denotes the temperature of the membrane 
and $\gamma$ denotes the hydrodynamic drag of the molecular species.
The fluctuations are given by $\mb{g}_{\tiny thm,q}$.  The $\Phi$ 
denotes the chemical potential for the molecular species at location $x$
given the protein is at location $\mb{X}$.  We consider the case 
here of a flat membrane and model the spatial fields of the 
system as having periodic boundary conditions in 
equation~\ref{equ_conc}.}

The membrane temperature field $\theta_C(x,t)$ is governed by
\begin{eqnarray}
\label{equ_temperature_mem}
\frac{\partial 	\theta_C(x,t)}{\partial t} & = & 
\frac{\nabla\cdot \left(\kappa_{CC}\nabla 	\theta_C(x,t)\right)}{c_C}
-\frac{\kappa_{CI}(x;X)
\left(	\theta_C(x,t) - \theta_I \right)}{c_C}  \\
\nonumber
&-&  \frac{c_0\nabla \Phi:\bar{\kappa}\nabla q(x,t) }{c_C}
+ \frac{\left(c_0\nabla \Phi\right):\frac{1}{\gamma}\left(c_0\nabla \Phi\right)}{c_C} 
+ \mb{g}_{\tiny thm, 	\theta_C}(x,t). 
\end{eqnarray}
\revB{The thermal conductivity of the membrane and interfacial region is
denoted by $\kappa_{CC}$, $\kappa_{CI}$ and the specific heat of the membrane
is denoted by $C_C$.  The fluctuations are given by $\mb{g}_{\tiny thm,
\theta_C}$.  We model the spatial fields of the system as having periodic
boundary conditions in equation~\ref{equ_temperature_mem}.}

\begin{figure}[!h]
\centerline{\includegraphics[width=0.95\columnwidth]{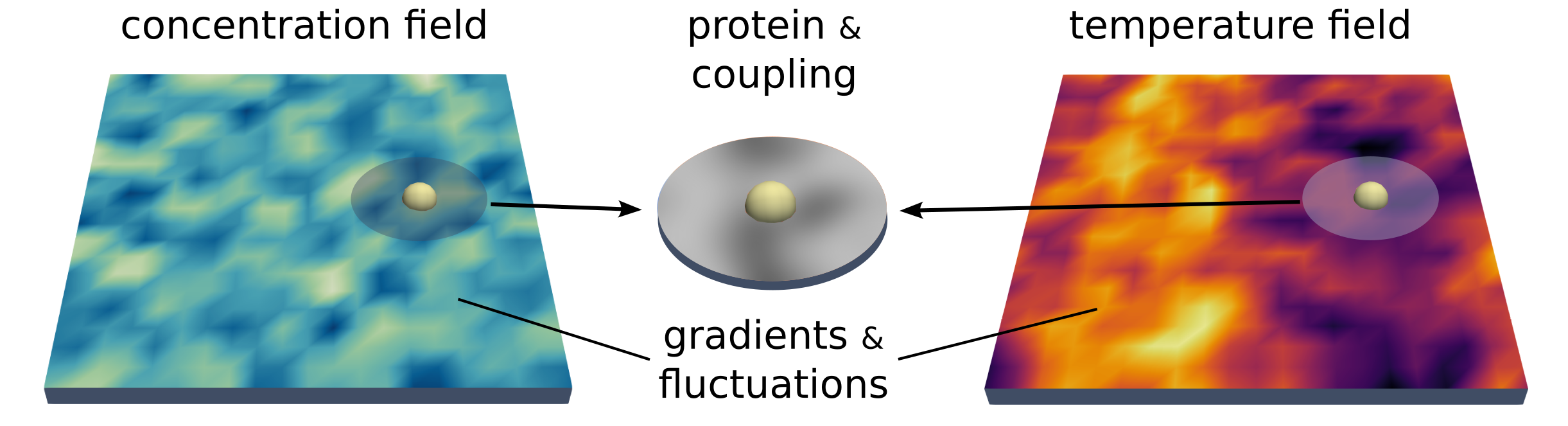}}
\caption{{\bf Membrane-Protein Drift-Diffusion: Stochastic Non-Equilibrium
Model.} \revB{Shows how the protein drift-diffusion dynamics are coupled to
surrounding concentration and temperature fields.  The interfacial region of
surrounding lipids is shown as a circular patch to denote the coarse-grained
range over which the coupling is active.  This is modeled through the term
$\kappa_{CI}(x,X)$ in equations~\ref{equ_temperature_mem}
and~\ref{equ_temperature_particle}.}
}
\label{fig_full_model}
\end{figure}

\revB{We model the nearby region of lipids surrounding the protein and its
surface as an interfacial region with temperature $\theta_I$.  This
models the coarse-grained coupling between a protein and 
lipids allowing for modeling effects that are under-resolved in
point-particle models.  This allows for modeling additional state information
for exchanges in energy and momentum  between the protein and the membrane.
We use in practice $\kappa_{CI}(x,X) = \kappa \eta(x - X)$ where $\eta(r)$ is a
radial function, such as a truncated Gaussian or Peskin
$\delta$-function~\cite{Atzberger2007,AtzbergerSoftMatter2016}.  We show the 
interfacial region and coupling in Figure~\ref{fig_full_model}.}

\revB{The other temperatures are governed by }
\begin{eqnarray}
\label{equ_temperature_particle}
\frac{\partial \theta_P}{\partial t} & = & -\frac{\kappa_{PI}\left(\theta_P -
\theta_I\right)}{c_P} 
+
\frac{\mb{F}_X^T \mb{M}_{\tiny XX} \mb{F}_X}{c_P} 
+ \mb{G}_{\tiny thm, \theta_P}, \\
\nonumber
\frac{\partial \theta_I}{\partial t} & = & \frac{\kappa_{PI}\left(\theta_P -
\theta_I\right)}{c_I} + \int \frac{\kappa_{CI}(x;X)\left( 	\theta_C(x,t) - \theta_I
\right)}{c_I} dx +
\mb{G}_{\tiny thm, \theta_I}.
\end{eqnarray}
\revB{The specific heats of the protein and interface are denoted 
by $C_P$, $C_I$.  The thermal conductivities are denoted by 
$\kappa_{PI}$, $\kappa_{CI}$.  The fluctuations
are given by $\mb{G}_{\tiny thm, \theta_P}$, $\mb{G}_{\tiny thm, \theta_I}$. 
}

\revB{We use a force acting on the protein that is coupled to the local 
concentration field given by}
\begin{eqnarray}
\mb{F}_X = -\partial_X U =  -\nabla_{\mb{X}} \Psi(\mb{X}) +
\int -\nabla_{\mb{X}} \Phi(x;\mb{X}) c_0 q(x,t) dx.
\end{eqnarray}
\revB{The $q(x,t)$ gives the fluctuating distribution of the concentration
field of the chemical species that interacts with the
protein~\cite{AtzbergerTabak2015,AtzbergerSoftMatter2016,AtzbergerRD2010}.  The
$\Psi(\mb{X})$ gives the potential energy of the protein.  The $\Phi(r,\mb{X})$
denotes a chemical potential for the free energy of the molecules of the
chemical species $q(x,t)$ to be at location $r$.   We treat the system in 
an over-damped regime where the hydrodynamic flow that would be 
induced by the drag of the dilute chemical species is negligible for the 
concentration field.  As seen in our model equations, the free energy that is 
associated with the force $\mb{F}_X$ and $-\nabla \Phi$  drives both the 
protein dynamics and fluxes of the concentration field of the chemical species. }

\rev{The system temperatures $\theta$ and related quantities are denoted using
notation of the form $\theta_{(\cdot)}$, with $\theta_P$ for the protein,
$\theta_C$ for the membrane, and $\theta_I$ for the interfacial region.  The
thermal conductivities for heat exchanges are given by $\kappa_{(\cdot)}$ with
$\kappa_{CC}$ for the membrane temperature field, $\kappa_{CI}$ between the
interface and membrane field, and $\kappa_{PI}$ between the protein and
interface.  In this work, we also simplify the models by treating
the hydrodynamic contributions of the membrane in the 
over-damped regime through the mobility tensor
$M_{XX}$ for the protein.  This helps to mitigate sources of stiffness in
the dynamics from the hydrodynamic relaxation time-scales.  For some phenomena,
this may yield results that differ from models that include the momentum 
of the hydrodynamic flows as in~\cite{tozzi2019out,
AtzbergerSurfFluctHydro2022,AtzbergerTabak2015}.
}

To account for the fluctuations we use Gaussian random fields $\mb{G}_{thm,*},
\mb{g}_{thm,*}$.  We discuss how to derive the specific form for these
stochastic driving fields in the next section.  We denote the vector-valued
terms as $\mb{G}_{thm}(t)$ and the stochastic spatial fields as $\gthm(x,t)$.  
To obtain these terms requires a non-equilibrium statistical mechanics analysis
of our system.
 
\subsection*{Non-equilibrium Statistical Mechanics of the Protein-Membrane System}
\revB{To provide a unified approach for analyzing the system
to derive the 
fluctuations and to develop our simulation methods, 
we also provide a more abstract reformulation of 
our model given in equations~\ref{equ_full_model}-~\ref{equ_temperature_particle}. }
We express our model using stochastic dynamics of the form 
\begin{eqnarray}
\label{equ_dY_gen}
\frac{d \mb{Y}}{d t} = \bar{K}^{(1)} \mathcal{D}\mathcal{S} + 
\bar{K}^{(2)} \mathcal{D}\mathcal{S} + 
\bar{K}^{(3)} \mathcal{D}\mathcal{S} + \mb{g}_{\tiny thm},  
\end{eqnarray}
Our formulation is based on the non-equilibrium statistical mechanics framework 
GENERIC of Ottinger in~\cite{Ottinger1997,hutter1998fluctuation}. 
We collect the system state variables together into the vector 
$\mb{Y}(t) = [\mb{X}(t),q(t),
\theta_P(t),\theta_C(t),\theta_I(t)]$. \revB{This allows for a 
uniform treatment of phenomena, including diffusivities that 
are dependent on location, temperature, or even friction that 
can have non-linear responses to velocity.  This reformulation also
provides insights helpful in analysis and derivations for obtaining 
the appropriate fluctuations and for
developing numerical approximations.  We perform analysis and
develop stochastic numerical methods that help ensure 
null-space alignment between the operators of the 
reversible and irreversible processes and preserve 
other properties of the dynamics.}

The energy of the system is denoted by $\mathcal{E}$, the entropy is
denote by $\mathcal{S} = S^{(1)} + S^{(2)} + S^{(3)}$.
The entropy $S^{(1)}$ is associated with the protein,
$S^{(2)}$ with the membrane, $S^{(3)}$ with the 
interfacial region.  The gradient in $\mb{Y}$ of the entropy 
is denote by 
$\mathcal{D}\mathcal{S}$.
\rev{The 
dissipative irreversible processes in the dynamics are modeled by the collection
of symmetric positive definite operators $\bar{K}^{(j)}$.  We give more details of
the entropy and these operators for our model below and 
in Appendix~\ref{appendix_K_j}. }

We model the fluctuations as
$\gthm =  \gthm^{(1)} + \gthm^{(2)} + \gthm^{(3)}$
with
\begin{eqnarray}
\label{equ_stoch_driving}
\gthm^{(j)} = k_B(\nabla \cdot K^{(j)}) + B^{(j)}\frac{dW_t^{(j)}}{dt},
\end{eqnarray}
where $B^{(j)}B^{(j),T}  =  2k_B K^{(j)}$.  The $dW_t^{(j)}$ are increments of
the Wiener process independent in $j$~\cite{Oksendal2000,Gardiner1985}.  The
$k_B$ is the Boltzmann constant~\cite{Reichl1997}.  We obtain the correlations
for the stochastic driving fields $\gthm^{(j)}$ by combining the increments
$dW_t^{(j)}$ using the operators $B^{(j)}$.  We use
equation~\ref{equ_stoch_driving} to obtain the fluctuations for our system,
from the dissipative operators $K^{(j)}$. We discuss these operators for our
model below.

For our model in equation~\ref{equ_full_model}--~\ref{equ_temperature_particle}, the energy is given by 
\begin{eqnarray}
\mathcal{E}(\mb{Y}) &=& \mathcal{E}^{(1)}(\mb{Y}) + \mathcal{E}^{(2)}(\mb{Y}) + \mathcal{E}^{(3)}(\mb{Y}).
\end{eqnarray}
The particle energy is given by 
$\mathcal{E}^{(1)}(\mb{Y}) = \Psi(\mb{X}) + c_P \theta_P.$
The potential energy is denoted by $\Psi(\mb{X})$.  
The energy of the concentration field depends on a 
spatial chemical potential $\Phi(x;\mb{X})$ with 
\begin{eqnarray}
\mathcal{E}^{(2)}(\mb{Y}) &=& \int \Phi(x;\mb{X}) c_0 q(x) dx + \int c_C \theta_C(x) dV_x.
\end{eqnarray}
The interface has the thermal energy 
$\mathcal{E}^{(3)}(\mb{Y}) = c_I \theta_I.$
We consider entropy arising from the particle, membrane, and interfacial coupling
\begin{eqnarray}
\mathcal{S}(\mb{Y}) & =& \mathcal{S}^{(1)} + \mathcal{S}^{(2)} + \mathcal{S}^{(3)}.
\end{eqnarray}
The entropy associated with the particle temperature $\theta_P$ is
\begin{eqnarray}
\mathcal{S}^{(1)} & = &  c_P \ln(\theta_P).
\end{eqnarray}
The membrane concentration field $q(x)$  and temperature $\theta_C(x)$ contribute the entropy
\begin{eqnarray}
\mathcal{S}^{(2)} & = & 
\mathcal{S}(q,\theta) = -\int c_0 q(x) \ln(q(x)) dx + \int c_C \ln(\theta_C(x)) dV_x. 
\end{eqnarray}
Entropy also arises from the temperature $\theta_I$ we track at the 
exchanges at the interface between the particle and the membrane 
\begin{eqnarray}
\mathcal{S}^{(3)} & = & c_I \ln(\theta_I).
\end{eqnarray}
We express the entropy gradient as
\begin{eqnarray}
\label{equ_grad_S}
\label{equ_grad_E}
\mathcal{D} \mathcal{S} 
&=& 
\left[
\partial_{\mb{X}} \mathcal{S},
\partial_{q(x)} \mathcal{S},
\partial_{\theta_P} \mathcal{S}, 
\partial_{\theta_I} \mathcal{S}, 
\partial_{\theta_C(x)} \mathcal{S} 
\right]^T.
\end{eqnarray}
This has the contributions
\begin{eqnarray}
\begin{array}{llllllll}
\partial_{\mb{X}} \mathcal{S} &=& 0, &
\partial_{q(x)} \mathcal{S} &=&
-c_0\left(1 + \ln(q(x))\right), && \\
\partial_{\theta_P} \mathcal{S} &=&
c_P/\theta_P,&
\partial_{\theta_I} \mathcal{S} &=& 
c_I/\theta_I,  \\
\partial_{\theta_C(x)} \mathcal{S} &=& 
c_C/\theta_C(x).
\end{array}
\end{eqnarray}
The energy gradient can be expressed as
\begin{eqnarray}
\label{equ_grad_E}
\mathcal{D} \mathcal{E} 
&=& 
\left[
\partial_{\mb{X}} \mathcal{E},
\partial_{q(x)} \mathcal{E},
\partial_{\theta_P} \mathcal{E}, 
\partial_{\theta_I} \mathcal{E}, 
\partial_{\theta_C(x)} \mathcal{E} 
\right]^T.
\end{eqnarray}
This has components
\begin{eqnarray}
\label{equ_total_energy_grad}
\begin{array}{llllllll}
\partial_{\mb{X}} \mathcal{E} &=& \nabla_{\mb{X}} \Psi(\mb{X}) + \int
\nabla_{\mb{X}} \Phi(x;\mb{X}) c_0 q(x) dx, &
\partial_{q(x)} \mathcal{E} &=& c_0\Phi(x;\mb{X}),  \\
\partial_{\theta_P} \mathcal{E} &=& c_P, & 
\partial_{\theta_I} \mathcal{E} &=& c_I, \\
\partial_{\theta_C(x)} \mathcal{E}&=& c_C.
\end{array}
\end{eqnarray}

\rev{This provides a statistical mechanics analysis of our model in equation 
~\ref{equ_full_model}
and a systematic way to derive the associated fluctuations and stochastic 
driving fields.  In particular, by expressing the dynamics in terms
of $\bar{K}^{(1)}$, $\bar{K}^{(2)}$, and $\bar{K}^{(3)}$ for 
equations~\ref{equ_full_model}--~\ref{equ_temperature_mem},  we can use 
equation~\ref{equ_stoch_driving} to  obtain the stochastic driving fields $\gthm^{(1)}$ , $\gthm^{(2)}$, and 
$\gthm^{(3)}$.  We give the operators $K^{(j)}$ for our model in equation~\ref{equ_full_model} in 
Appendix~\ref{appendix_K_j}.  This provides the form of the fluctuation terms for the
membrane-protein system dynamics in 
equations~\ref{equ_full_model}--~\ref{equ_temperature_mem}.
For performing practical simulations of the protein-membrane system, 
numerical methods are required to discretize the equations and 
to generate efficiently the samples of the stochastic terms $\gthm^{(j)}$. }

\subsection*{Stochastic Numerical Methods for Simulations of the Membrane-Protein System} 
\label{sec_stoch_num}

\rev{We now discuss briefly our simulation approaches and  
stochastic numerical methods for capturing the discrete
particle drift-diffusion dynamics and the fluctuations of the 
continuum concentration and temperature fields.
We remark that we focus in this paper primarily on 
the biophysical motivations of the work 
and will discuss further technical details
of the developed numerical methods elsewhere.
\revB{We provide further details on the methods in 
Appendix~\ref{appendix_R_j} and numerical 
validation studies for convergence in 
Appendix~\ref{appendix_validation}.}
}

\subsubsection*{Temporal Discretization and Time-Step Integration}
\revB{We discretize and integrate the stochastic dynamics in 
equation~\ref{equ_dY_gen} using the following two-stage approach}
\begin{eqnarray}
\label{eqn_stoch_num_method}
\tilde{Y}^{n+1} &= & Y^n + a(Y^n) \Delta{t} + \sum_j 
b^{(j)}(Y^n)\Delta{W}^{n,j} \\ 
\nonumber 
Y^{n+1} & = & Y^n 
+ \frac{1}{2} \left(a(Y^n) + a(\tilde{Y}^{n+1})\right)\Delta{t}
+ \frac{1}{2} \sum_j \left(b^{(j)}(Y^n) + b^{(j)}(\tilde{Y}^{n+1})\right)\Delta{W}^{n,(j)}.
\end{eqnarray}
The $a(Y) = L(Y)\nabla E(Y) + \sum_j \nabla K^{(j)}(Y) S^{(j)}(Y)$
and $b^{(j)}(Y) = B^{(j)}(Y)$.  The Wiener increments $\Delta{W}^{n,(j)}$ 
denote Gaussian random variates having mean zero and 
variance $\langle \Delta{W}^{n_1,j_1} \Delta{W}^{n_2,j_2} \rangle 
= \delta_{n_1,n_2} \delta_{j_1,j_2} \Delta{t}$.
\revB{Our integrator is based on a variant of the 
Euler-Heun Method~\cite{Platen1992}.  There also have been alternative integrators 
developed for related non-equilibrium formulations
in~\cite{hutter1998fluctuation}.}
An important part of the updates is that the same increments 
$\Delta{W}^{n,j}$ are used in both steps.  This serves as part of how 
the contributions of the divergence term are handled implicitly by 
the numerical methods~\cite{Platen1992}.
We remark that the updates are equivalent to approximating
the Stratonovich formulation of the stochastic process where fluctuations are
treated as 
$g_{thm}^{(j)} = B^{(j)}\circ dW_t^{(j)}$.  When giving the process the 
Ito interpretation the $\circ$ operator would expand the expression 
to include the drift divergence term as above.  The $B^{(j)}$ satisfy
$B^{(j)}B^{(j),T}  =  2k_B K^{(j)}$ with the Boltzmann constant $k_B$
and the operators $K^{(j)}$ given above.

\subsubsection*{Spatial Discretization of the Continuum Fields}
\begin{figure}[!h]
\centerline{\includegraphics[width=0.95\columnwidth]{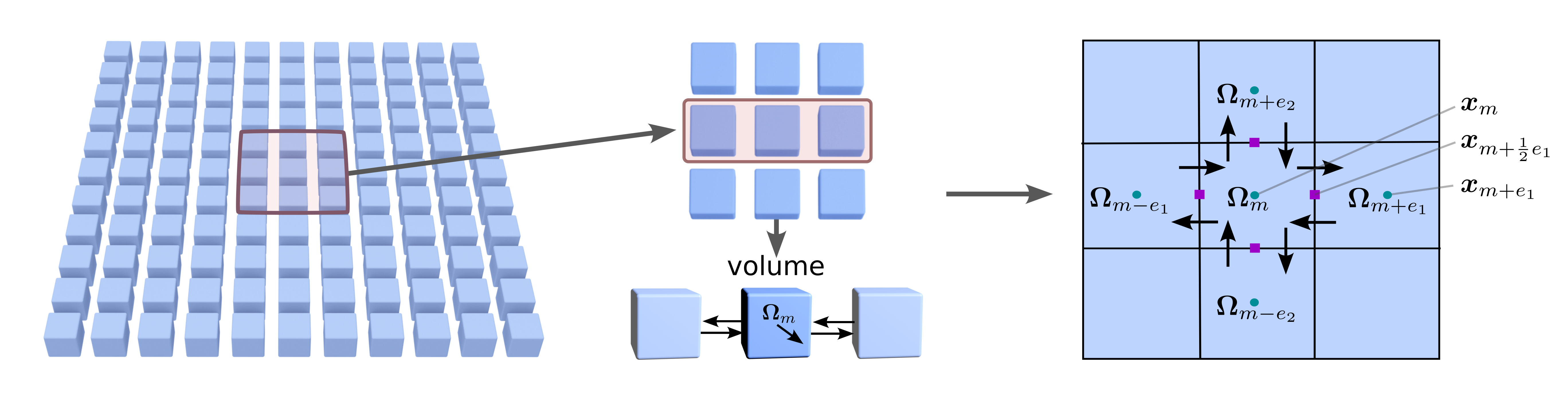}}
\caption{\rev{{\bf Spatial Discretization.}
The system is spatially discretized using a finite volume approach where
continuum fields on $\Omega = \cup_m \Omega_m$ are divided into a finite
collection of volumes $\Omega_m$.  The gradients and divergences are
approximated by discrete operators $\mathcal{G}$ and $\mathcal{D}$ modeling the
fluxes and exchanges between the volumes.  This ensures the stochastic
numerical methods adhere to physical conservation and adjoint conditions.}
}
\label{fig_spatial_discr}
\end{figure}

\rev{We spatially discretize the system using a finite volume approach.  This is
done by providing discretizations for the divergence 
$\mathcal{D}$ and gradient $\mathcal{G}$.  
We generate numerical methods for the spatial discretization by
replacing throughout in our analytic expressions $\nabla$ by 
the discrete operator $\mathcal{G}$ 
and $\nabla \cdot = \mbox{div}$ by discrete operator $\mathcal{D}$.
\revB{We build on our 
finite volume methods in~\cite{Atzberger2007,AtzbergerRD2010}.
Recently, there also has also been work on discretization 
methods using finite elements and other discretizations to
generate fluctuations with $O(N)$ complexity 
in~\cite{martinez2024finite,de2016finite}.
}
In our finite volume methods, we replace spatial integrals 
$\int (\cdot) dx$ by the corresponding finite sums 
$\sum_{m} (\cdot)_{x_m} \dvol$.  We treat continuum bodies as 
divided into a discrete finite collection of boxes each having 
volume $\dvol$.  
The $x_m$ denotes the location
of the $m^{th}$ finite volume box, see Figure~\ref{fig_spatial_discr}. }

We consider fields represented by average values at the volume 
centers $\mb{x}_m$.  We discretize the operators 
by considering fluxes $\mb{J}$ 
at location of the volume boundaries $\mb{x}_{m\pm\frac{1}{2}e_d}$,
using related conventions as in our 
finite volume methods in~\cite{Atzberger2007,AtzbergerRD2010}.  
The $e_d$ denotes the standard basis vector with 
zeros for all entries except for the value $1$ 
for entry with index $d$.  \revB{This is achieved by 
discretizing each of the gradient operators 
$\mathcal{G} = \mbox{grad}(\cdot)$ of the field values $\mb{F}$ with 
components $F^{(d)}$
by using the central differences}
\begin{eqnarray}
\label{equ_fd_grad}
\mbox{grad}(\mb{F})^{(d)}(\mb{x}_{m \pm \frac{1}{2}e_d})
= \mathcal{G}(\mb{F})^{(d)}(\mb{x}_{m \pm \frac{1}{2}e_d})
= 
\pm\frac{1}{\Delta{x}} 
\left(
{F}^{(d)}(\mb{x}_{m \pm e_d})
-
{F}^{(d)}(\mb{x}_{m})
\right).
\end{eqnarray}
The fluxes are given by 
$\mb{J}^{(d)}(\mb{x}_{m \pm \frac{1}{2}e_d}) 
= \mbox{grad}(\mb{F})^{(d)}(\mb{x}_{m \pm \frac{1}{2}e_d})$
We discretize the 
divergence operators $\mathcal{D} = \mbox{div}(\cdot)$ of
the fluxes by the central differences 
\begin{eqnarray}
\label{equ_fd_div}
\mbox{div}(\mb{J})(\mb{x}_m) = \mathcal{D}(\mb{J})(\mb{x}_m) 
= \frac{1}{\Delta{x}} \sum_{d = 1}^{n} 
\left\lbrack
\mb{J}^{(d)}(\mb{x}_{m + \frac{1}{2}e_d})
- \mb{J}^{(d)}(\mb{x}_{m - \frac{1}{2}e_d})
\right\rbrack.
\end{eqnarray}
This corresponds to discretizing within the operator $K^{(j)}$ the gradient and
divergence operators using $\mathcal{G}$ and $\mathcal{D}$ above in
equation~\ref{equ_full_model}--~\ref{equ_temperature_particle}. We further have
that the discretized gradient operator $\mathcal{G}$ appearing in these
expressions is the negative adjoint of the discretized divergence operator
$\mathcal{D}$, so $\mathcal{G} = -\mathcal{D}^T$.  
Our finite volume approach provides a
systematic way to obtain discretizations and stochastic numerical methods 
that satisfy properties such as the adjoint relations between gradient
and divergence which help preserve structural features of the dynamics 
important in their statistical mechanics~\cite{AtzbergerRD2010,AtzbergerSELM2011}.
This provides numerical methods for handling the spatial 
discretization and time-step integration of our model in 
equation~\ref{equ_full_model}--~\ref{equ_temperature_particle}.  

\subsubsection*{Methods for Generating Fluctuations} 
To obtain practical simulation methods, we need to handle
the fluctuations of the continuum concentration and temperature fields of
the membrane given by $\gthm^{(j)}$ in equation~\ref{equ_stoch_driving}.
This requires being able to generate efficiently the stochastic driving 
fields each time-step $\mb{h}_{thm}^{(j),n} = B^{(j)}(Y)\Delta{W}^{n,j}$, where 
$B^{(j)}B^{(j),T}  =  2k_B K^{(j)}$.  Methods such as Cholesky 
Factorizations are prohibitive for the continuum fields given that they scale as
$O(N^3)$ in the number of degrees of freedom
$N$~\cite{AtzbergerSELM2011,Trefethen1997}.  
Further, the operator $B^{(j)} = B^{(j)}(\mb{Y}^n)$ in general depends on the state $\mb{Y}^n$
which would require recomputing these factors each time-step as the state changes. 
We show alternatives can be developed avoiding these issues
through a combination of analytic factorizations and 
further reductions.  
We generate the fields using the formulation 
\begin{eqnarray}
\label{equ_h_thm}
\mb{h}_{thm}^{(j)} = \sqrt{\Delta{t}}B^{(j)} \bsy{\xi}^{(j)} =  \sqrt{2k_B\Delta{t}} R^{(j)} \bsy{\xi}^{(j)},
\end{eqnarray}
where the $\bsy{\xi}^{(j)} \sim \eta(0,I)$ are standard Gaussian random
variates and $R^{(j)}$ is a factor satisfying $K^{(j)} = R^{(j)}R^{(j),T}$. 
\rev{We perform analysis to find explicit expressions for the factors $R^{(j)}$
for the proteins, interface, and membrane in Appendix~\ref{appendix_R_j}.  
This
allows us to generate the needed stochastic fields efficiently each time step
needed in equation~\ref{equ_h_thm} and~\ref{equ_stoch_driving}.  We decompose
the operators into parts allowing for generation with sampling methods having
computational complexity at most $O(N^2)$.  For some terms we are able to
obtain better results by using sparsity and other structures to achieve
algorithms having sampling complexity $O(N)$.  \revB{We give further details on our
analytic factorizations and discussion of our stochastic sampling methods in
Appendix~\ref{appendix_R_j} and Appendix~\ref{appendix_validation}.}}

\section*{Results}

\rev{We investigate heterogeneous membranes and the 
impacts of spatial variations in concentration
and temperature on the drift-diffusion dynamics of 
proteins.  We consider 
(i) how concentration gradients and kinetics 
can drive the spatial organization of proteins, 
(ii) the roles of fluctuations in the encoding
of signals by proteins to sense  
external thermal gradients, and (iii) how localized 
laser heating can be used to 
probe protein escape kinetics from local 
energy wells within heterogeneous membranes.  The results
demonstrate a few ways the simulation approaches can be 
used to study biological mechanisms within membranes 
and related non-equilibrium phenomena in other systems.}

\subsection*{Protein Drift-Diffusion in Concentration Gradients of Heterogeneous 
Membranes}

\begin{figure}[!h]
\centerline{\includegraphics[width=0.99\columnwidth]{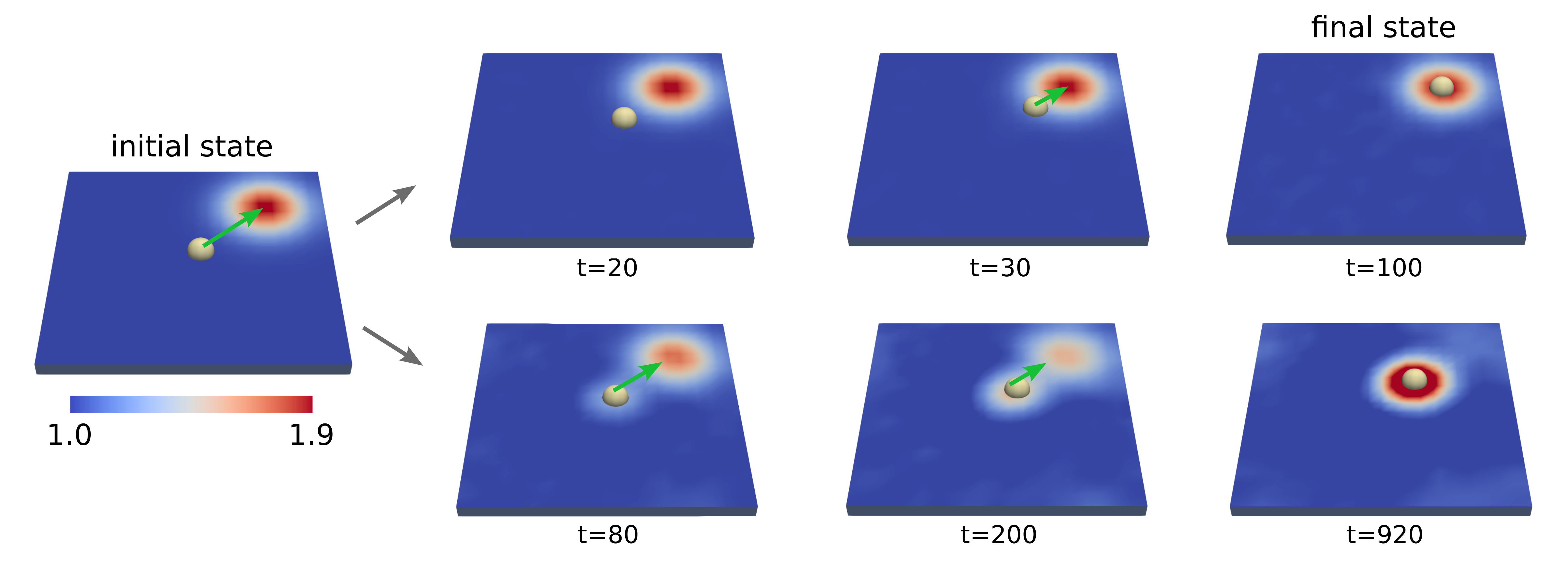}}
\caption{\revB{{\bf Protein Positioning through Concentration of Signaling Molecules.} 
We show how an individual protein interacts with the concentration field of signaling
molecules with the affinity given in equation~\ref{equ_conc_energy}.  We show
the evolution over time $t$ the number of time-steps with the parameters
in Table~\ref{table_conc_grad}.  Depending on
the relative time-scales of the signaling molecule diffusivity and protein
dynamic time-scale, there are different behaviors.  When the signaling molecule diffusivity
is small, the protein localizes to the initial concentration location \textit{(top)}.  
When the signaling molecule diffusivity is larger, the protein does not move significantly
and the concentration collects around the initial location of the 
protein \textit{(bottom)}.}
}
\label{fig_conc_grad_schematic}
\end{figure}

Protein organization can be governed by other cellular metabolic activities
that create regions with enhanced concentrations of chemical 
species~\cite{goni2014basic,nicolson2023fluid,singer1972fluid}.  
We model the concentration of a signaling chemical species by the field  
$q(x)$.  We consider the case where there is an interaction energy of the form 
\begin{eqnarray}
\label{equ_conc_energy}
\mathcal{V}(\mb{X};q) = \int \eta(x - \mb{X}) c_0 q(x) dx + \Psi(\mb{X}).
\end{eqnarray}
This has the associated force
\begin{eqnarray}
\mb{F}_{\scriptstyle X} = -\nabla_X \mathcal{V}(\mb{X};q) = \int \nabla_x \eta(x - \mb{X})
c_0q(x) dx -\nabla_X \Psi(\mb{X}).
\end{eqnarray}
We use for the coupling kernel
\begin{eqnarray}
\label{equ_eta_conc}
\eta(|s|) & = & \frac{k_1}{Z} \exp \left( \frac{-|s|^2}{2\sigma_0^2} \right),
\hspace{1cm} Z = \left( 2\pi \sigma_0^2 \right)^{d/2}.
\end{eqnarray}
For the membrane, we take $d = 2$, $k_1 = 1.1$ for the coupling strength,
and $\sigma_0 = 0.2$.  \revB{The energy in equation~\ref{equ_conc_energy} is
motivated by the free-energy associated with interactions between the signaling
molecules and the protein, such as through electrostatics or other physical 
effects.  As a basic model we use the coupling kernel in equation~\ref{equ_eta_conc}
to give a model for a localized decaying interaction.}
 
As an initial model for a single particle, we use for simplicity the mobility
$\mb{M}_{\tiny XX} = ({1}/{\gamma_p})\mathcal{I}$.  The $\gamma_p$ denotes
the effective hydrodynamic drag of the particle for translational motions
within the membrane.  \revB{More sophisticated models for multiple particles can also
be formulated using the Saffman-Delbruck theory and other hydrodynamic approaches,
see \cite{AtzbergerSoftMatter2016,
AtzbergerSurfFluctHydro2022,Saffman1976}.
Using these models our approaches readily can be extended to 
take into account the hydrodynamic interactions mediated by 
flow of the surrounding lipids of the membrane and the solvent.
}

\begin{figure}[h]
\centerline{\includegraphics[width=0.6\columnwidth]{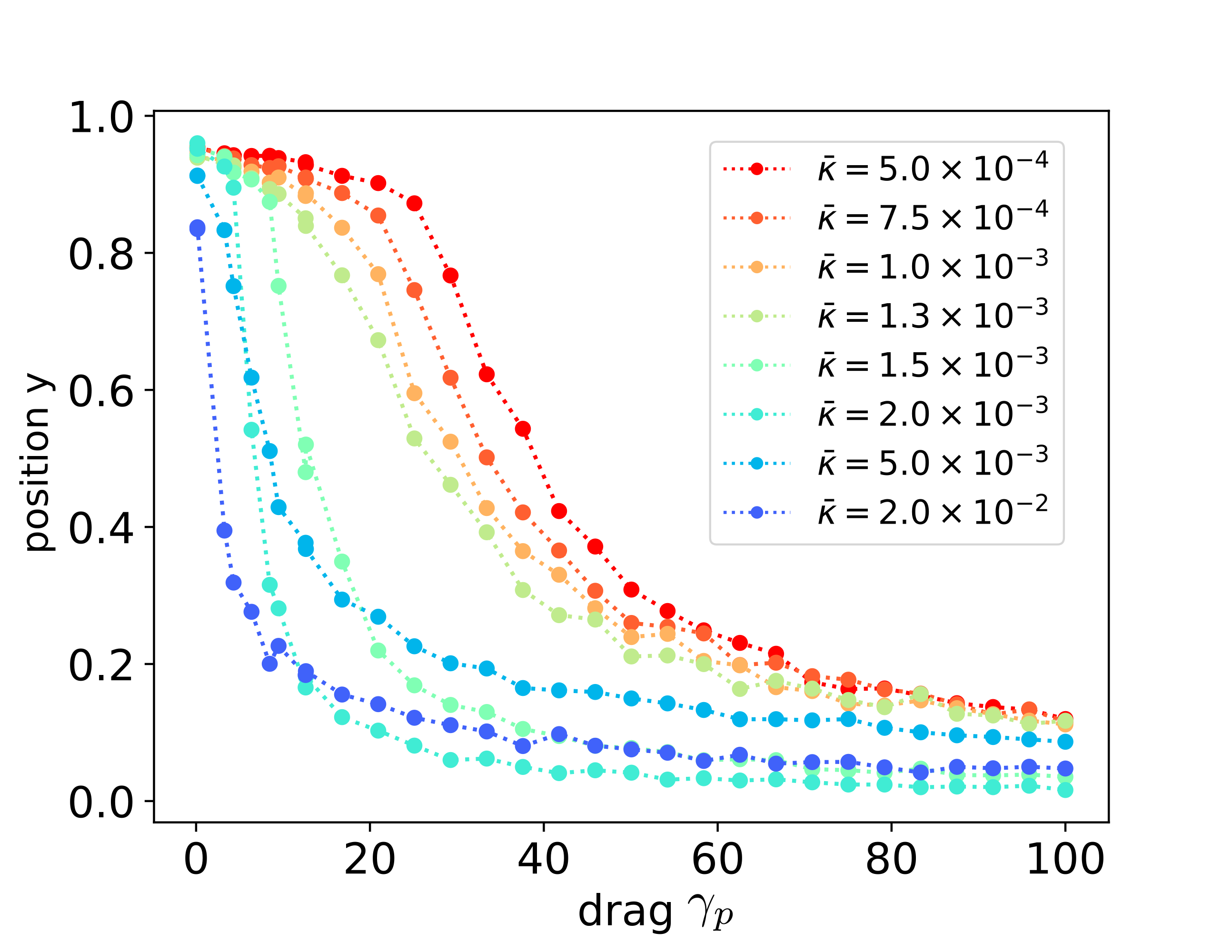}}
\caption{\revB{{\bf Protein Positioning Through Concentration of Signaling Molecules.} 
We show how key factors impact the positioning of proteins in 
response to attraction to a region of large concentration of a signaling
molecule.  The scaled final protein position
$y = (x^{(2)} - x_0^{(2)})/(x_1^{(2)} - x_0^{(2)})$ is shown 
in range $0.0$ and $1.0$.  This gives the 
fraction of the distance it moved toward the
signaling molecule initial source location $x_1$.  
We consider cases when varying the
signaling molecule diffusivities $\bar{\kappa}$ and the effective hydrodynamic
radius of the protein characterized by the viscous drag $\gamma_p$.  The
localization of the protein involves a competition between the diffusion of the
signaling molecules and protein motion toward regions of larger concentration.
We also show a few example cases in
Figure~\ref{fig_conc_grad_schematic}. }
}
\label{fig_conc_grad_results}
\end{figure}

In our simulation studies, we consider the case where there is initial
concentration of a signaling chemical species in a localized region
centered at $x_1$.  This is modeled by a Gaussian concentration with 
mean $\mu_0 = x_1 = [1.5,1.5]$ and variance $\sigma^2$ with $\sigma = 0.2$.  
We then consider how the simultaneous concentration drift-diffusion 
of $q(x,t)$ over
time interacts with the particle drift-diffusion $\mb{X}(t)$ when 
$\mb{X}(0) = x_0 = [0,0]$.  Given the
affinity between the particle and signaling chemical species, we see over time
they will occupy the same location through mutual attraction, see
Figure~\ref{fig_conc_grad_schematic}.  On much longer time-scales the particle
and signaling chemical species can also diffuse together throughout the domain.
We show a few cases for how the particle and concentration field evolve in
Figure~\ref{fig_conc_grad_schematic}. 

We investigate the localization of the protein and signaling species
concentration field when varying the protein drag $\gamma_p$ and the signaling
species diffusivity $\bar{\kappa}$.  We performed simulations of the system
using the parameters given in Table~\ref{table_conc_grad}.  We show results in
Figure~\ref{fig_conc_grad_results}.

\begin{table}[!h]
\setlength{\tabcolsep}{4pt} 
\global\long\def\arraystretch{1.5}
\centering {\fontsize{7}{8}\selectfont 
\begin{tabular}{|l|l|l|l|l|l|}
\hline 
\rowcolor{atz_table1} \multicolumn{2}{|l|}{\textbf{parameter}} & \textbf{value} & \multicolumn{2}{l|}{\textbf{parameter}} & \textbf{value} \tabularnewline
\hline 
$\bar{\kappa}$ & concentration diffusion & \atznum{1.2e-3} & $C_{P}$ & specific heat: particle & \atznum{1.2}\tabularnewline
\hline 
$c_{0}$ & total concentration & \atznum{2.1} & $C_{C}$ & specific heat: membrane & \atznum{1.3e2}\tabularnewline
\hline 
$\kappa_{PI}$ & heat conduction: particle & \atznum{8.2e6} & $C_{I}$ & specific heat: interface & \atznum{1.4e2}\tabularnewline
\hline 
$\kappa_{CI}$ & heat conduction: interface & \atznum{3.02e3} & $\gamma_{p}$ & particle drag & \atznum{12.6}\tabularnewline
\hline 
$\kappa_{CC}$ & heat conduction: membrane & \atznum{1.3e2} & $\theta_{0}$ & baseline membrane temperature & \atznum{3.0}\tabularnewline
\hline 
$\kappa_{0}$ & heat conduction: fluid & \atznum{2.1e-3} & $k_{B}$ & Boltzmann's constant & \atznum{1e-5}\tabularnewline
\hline 
$n_{x}$ & number grid points in x & \atznum{20} & $\Delta x$ & mesh spacing & \atznum{0.1}\tabularnewline
\hline 
$n_{y}$ & number grid points in y & \atznum{20} & $\Delta t$ & time step & \atznum{1e-3}\tabularnewline
\hline 
\end{tabular}\vspace{0.2cm}
 }\caption{\rev{\textbf{Parameters for the Concentration Gradient Model.} We give the values
for the SELM simulations of the signaling molecule concentration
fields and particle drift-diffusion dynamics.}}
\label{table_conc_grad}
\end{table}

\revB{We find that varying the signaling molecule diffusivity $\bar{\kappa} =
\theta_C/\gamma$ and the protein drag $\gamma_p$ can be used to regulate
localization. 
The final location of the protein depends on the ratio of the
time-scale $\tau_s = \ell^2/\bar{\kappa}$ for the signaling 
chemical species to diffuse and the time-scale $\tau_p = \ell \gamma_p/f_0$ for 
the protein motion.  The $\ell$ is the initial separation distance and $f_0$
the strength of the initial interaction force on the protein.  
If $\tau_s$ is small relative to $\tau_p$, then we find the
protein moves to location $x_1$ of the initial large concentration.  If the 
$\tau_s$ is large relative to $\tau_p$, we find the signaling species 
migrates to surround the protein at location $x_0$ 
before the protein has the chance to move significantly.}
 
\revB{At intermediate time-scales we find there is a combination of these effects 
with both the signaling molecules and protein meeting at a location in-between
the locations $x_0$ and $x_1$.}  Interestingly, as the diffusivity of the signaling
molecule becomes very large there can be some reversals since it can accumulate
rapidly as it moves all at once toward the protein which results in a large 
localized force that briefly pulls the protein toward the location $x_1$.  
We show the results of our studies in Figure~\ref{fig_conc_grad_results}.

\subsection*{Thermal Gradient Sensing and Fluctuations}

\begin{figure}[h]
\centerline{\includegraphics[width=0.99\columnwidth]{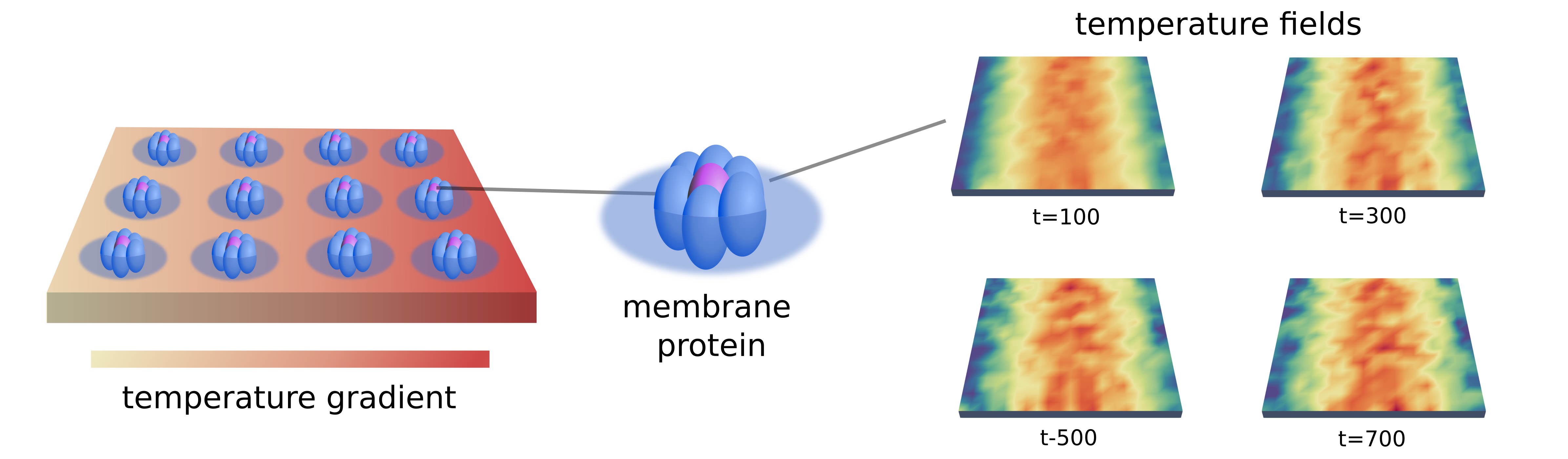}}
\caption{\rev{{\bf Sensing of Thermal Gradients.}  We consider the responses of
thermal sensitive proteins and how they may encode information about spatial
temperature variations obscured by fluctuations \textit{(left)}.  We show SELM
simulations of fluctuating temperature fields having an initial periodic
structure given by $\theta_C(\mb{x},0) = \theta_0\left(1.0 + a_0\sin(\pi \mb{k}\cdot
\mb{x}/L)\right)$ with $\mb{k} = [1,0]$ and amplitude $a_0 = 1.0$, \textit{(right)}.}
}
\label{fig_thermal_grad_schematic}
\end{figure}

The detection of changes in temperature and thermal gradients plays an
important role in many types of cells.  This includes intracellular processes
involved in modulating
growth~\cite{ye2021molecular,oyama2015directional,chuma2024implication} and for
single cell micro-organisms the ability to control migration toward or away
from heat sources~\cite{okabe2021intracellular,
zhu2025exploring,jiang2009mechanism}.  
We consider thermal sensitive proteins, such as the channel proteins TRP that
have temperature dependent gating
dynamics~\cite{caterina1997capsaicin,zhang2015molecular,vlachova2024human}
for detecting 
changes~\cite{sengupta2013sensing,jiang2009mechanism,ye2021molecular}.  In our
modeling, these are patterned within the membrane at fixed locations and we
investigate how they could encode spatial variations of the temperature when
obscured by fluctuations.  For transmission of heat to the 
local protein from an area of the surrounding temperature field, we use 
\begin{eqnarray}
\bar{Q}(\mb{x}_i,t) = \int \zeta(\mb{x} - \mb{x}_i) \theta_C(\mb{x},t) dx, \;\;
\zeta(|s|)  =  \frac{1}{Z} \exp \left( \frac{-|s|^2}{2\sigma_0^2} \right),
\;\; Z = \left( 2\pi \sigma_0^2 \right)^{d/2}.
\end{eqnarray}
\revB{This $\bar{Q}$ models the temperature that would be sensed by a fixed 
protein, such as a TRP channel at location $\mb{x}_i$ within the 
membrane~\cite{zhang2015molecular}. 
We use parameters $d = 2$ and $\sigma_0 = 0.1$.}

\revB{We investigate the fidelity by which the proteins can 
collectively be used to sense
spatial variations in the temperature in the presence of fluctuations.
Since TRP channel gating gives a permeability to ions that depends on  
the local temperature, this can be used in chemical reactions to increase
the local concentration of activated protein molecules.  
In this way the spatial variations in the 
concentration of the activated protein species $I(\mb{x}_i,t)$ can be used to
locally encode the external temperature gradient for use in further 
downstream chemical reactions within the cell.  We can consider 
reactions similar to those that arise in chemotaxis, such as the 
models in~\cite{AtzbergerRD2010,wang2011signaling}.}  

\revB{For chemical reactions with small spatial diffusion having motifs 
that arise in chemotaxis, the final concentrations 
that rapidly reach their equilibrium can be obtained by a reduction to 
a time-averaged response function
of the input signal~\cite{AtzbergerRD2010,alon2007network,pahlajani2011stochastic}.  
The equilibration of the chemical reactions serves to produce a signaling filter
for the gradient and fluctuations.  As a model for such a reduction, we use 
an effective time-averaged signal encoded in the local
concentration $\bar{I}$.  We use the averaging }
\begin{eqnarray}
\bar{I}(\mb{x}_i,t) = \beta_0 \int_{-\infty}^t 
\eta(t - s) \bar{Q}(\mb{x}_i,s) ds.
\end{eqnarray}
\revB{The $\eta$ gives the response function given the local reaction chemistry.
Motivated by first-order chemical 
reactions for activation and deactivation of a protein species for encoding,
we consider exponential response functions.  In particular, 
we use the response function $\eta(\tau) = \lambda
\exp(-\lambda\tau)$ with $\lambda = 10^4$, $\beta_0 = 1/3$. 
Other chemical reaction motifs would correspond to
different choices of the response function
$\eta$~\cite{jin2013gradient,AtzbergerRD2010,wang2011signaling,alon2007network}. } 

\revB{The $\bar{I}$
gives the concentration of the signaling molecules that encode the
spatially-temporally filtered temperature signal.  The signaling concentrations
$\bar{I}(\mb{x}_i)$ can be further coupled to downstream chemical reactions
that impact cellular
processes~\cite{jin2013gradient,wang2011signaling,AtzbergerRD2010}.  We focus
here on this initial processing of signals from the surrounding 
fluctuating temperature fields.}

\begin{figure}[h]
\centerline{\includegraphics[width=0.6\columnwidth]{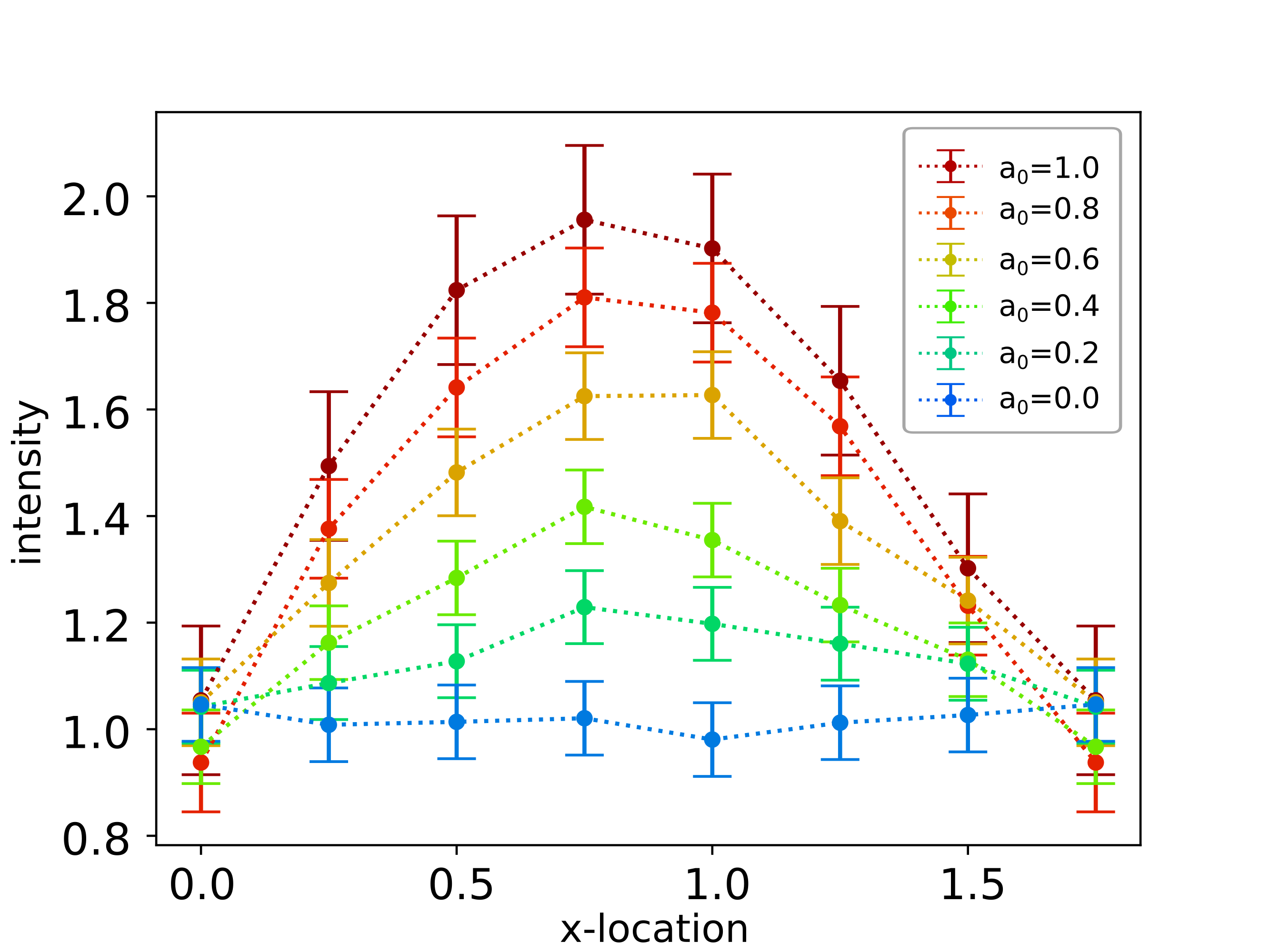}}
\caption{\rev{{\bf Sensing Thermal Gradients.}  We show results for protein
responses for encoding signals from a spatially varying temperature field
subject to fluctuations.  Shown is the average intensity of the indicator 
$\bar{I}$ concentration
and one standard deviation as error bars.  We investigate how the sensed signals 
change as the amplitude $a_0$ of the spatial temperature fields are varied.}
}
\label{fig_thermal_grad_results}
\end{figure}
We investigate the thermal sensing as the amplitude of the 
spatial variations of the external temperature field is varied.
We consider the case of spatial variations that start from
an initial temperature profile 
$\theta_C(\mb{x}) = \theta_0\left(1 + a_0\sin(\pi \mb{k}\cdot\mb{x}/L)\right)$,
with $\mb{k} = [1,0]$ and $a_0 = 1.0$.  The parameters of our model and 
simulation studies are given in Table~\ref{table_energy_well}.  
We show results in Figure~\ref{fig_thermal_grad_results}.

\begin{table}[!h]
\setlength{\tabcolsep}{4pt} 
\global\long\def\arraystretch{1.5}
 \centering {\fontsize{7}{8}\selectfont 
\begin{tabular}{|l|l|l|l|l|l|}
\hline 
\rowcolor{atz_table1} \multicolumn{2}{|l|}{\textbf{parameter}} & \textbf{value}
& \multicolumn{2}{l|}{\textbf{parameter}} & \textbf{value} \tabularnewline
\hline 
$\kappa_{PI}$ & heat conduction: particle & \atznum{8.2e6} & $C_{P}$ & specific heat: particle & \atznum{1.0}\tabularnewline
\hline 
$\kappa_{CI}$ & heat conduction: interface & \atznum{0.0} & $C_{C}$ & specific heat: concentration & \atznum{4e1}\tabularnewline
\hline 
$\kappa_{CC}$ & heat conduction: membrane & \atznum{8.2e4} & $C_{I}$ & specific heat: interface & \atznum{1.4e2}\tabularnewline
\hline 
$\kappa_{0}$ & heat conduction: fluid & \atznum{8.2e4} & $\theta_{0}$ & baseline membrane temperature & \atznum{3.0}\tabularnewline
\hline 
$c_{0}$ & total concentration & \atznum{2.1} & $k_{B}$ & Boltzmann's constant & \atznum{1e-3}\tabularnewline
\hline 
$n_{x}$ & number grid points in x & \atznum{20} & $\Delta x$ & mesh spacing & \atznum{0.1}\tabularnewline
\hline 
$n_{y}$ & number grid points in y & \atznum{20} & $\Delta t$ & time step & \atznum{1e-5}\tabularnewline
\hline 
\end{tabular}\vspace{0.2cm}
 } \caption{\rev{\textbf{Parameters for the Temperature Sensing Model.} We give the values 
for the SELM simulations of protein sensing of fluctuating temperature 
variations.}}
\label{table_energy_well}
\end{table}

We find while fluctuations can obscure significantly temperature gradients on
small spatial-temporal scales this can be mitigated by processes that 
serve to filter the signal.  In
our simulations the temperature fields start with an initial sinusoidal profile
and evolve over time toward a uniform
equilibrium while also undergoing spontaneous fluctuations from transient local
energy exchanges.  We can see that for the largest amplitudes the signal
of temperature changes can be detected, but 
becomes suppressed by the filtering over time and space as seen in the
indicator species responses $\bar{I}$.  For the smallest amplitudes, we 
see the gradient becomes obscured by noise. 
These simulations indicate some of the interesting trade-offs
between the level of filtering to obtain a reliable signal while still
resolving the spatial and temporal information inherent in the surrounding
temperature fields relevant for biological responses, 
see Figure~\ref{fig_thermal_grad_results}.    This gives 
some demonstrations of how the non-equilibrium SELM simulation approaches 
can be utilized to investigate biophysical signal transduction of temperature
gradients.

\subsection*{Hot Brownian Motion of Particles in Temperature Gradients}

\rev{We consider the non-equilibrium diffusion of particles that can undergo
temperature changes from environmental and external heating.  These
thermal effects can drive more rapid particle diffusion and
other phenomena referred to as 
Hot Brownian Motion~\cite{rings2010hot,schachoff2015hot}.  
Recent theoretical and simulation work studying these effects
include~\cite{schachoff2015hot,chakraborty2011generalised,
rings2010hot,riedel2015heat}.  In these studies,  
a description of the Brownian motion of a particle 
subject to laser heating is developed where temperature differences 
augment the local viscosity and fluctuations.  Related experiments were 
also performed by laser heating gold nanoparticles that 
exhibit significant variations in their diffusion~\cite{schachoff2015hot}. 
In other experimental observations,
heating caused locally by catalytic enzymatic reactions also were found to impact 
diffusion \cite{riedel2015heat}.}

The current theoretical studies reduce descriptions to particle-based models
that use a separation of time-scales between the changes in the particle temperature
and the spatial changes in the surrounding temperature fields of the environment.
For systems that have more persistence or externally imposed spatial gradients, 
we show how our SELM modeling and simulation approaches can be used to capture 
further spatial-temporal effects.  For example, the impacts of spatial heterogeneity, 
energy transfer and other augmentations from past locations that hot particles visit, 
or other time-scales associated with the surrounding environment.

\begin{figure}[!h]
\centerline{\includegraphics[width=0.99\columnwidth]{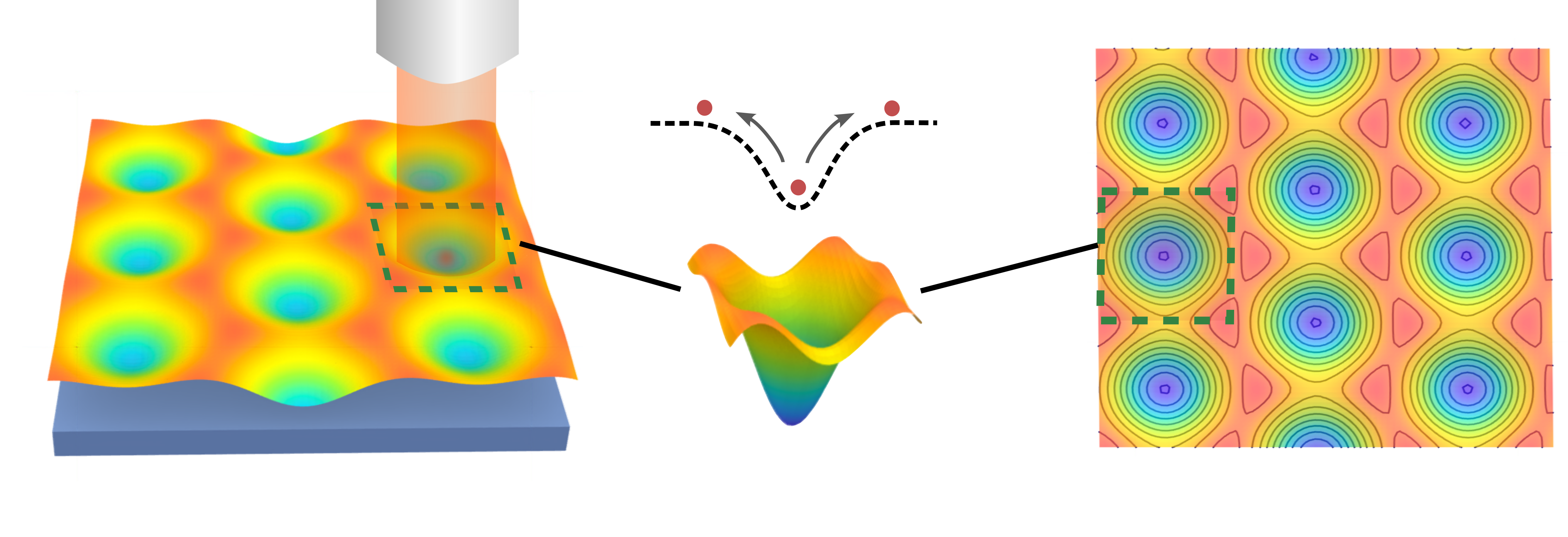}}
\caption{\rev{{\bf Hot Brownian Motion in Energy Wells.}  We consider particles 
undergoing Brownian motions which can change temperature from 
energy exchanges with the surrounding environment.  We study diffusion within 
a heterogeneous membrane where there are local energy wells some of which 
are heated by an external source.    
This impacts the particle diffusion within such wells and the 
time to escape by overcoming the energy barriers.  
We show results for particle escape times in 
Figure~\ref{fig_hot_brownian_results}.
}}
\label{fig_hot_brownian_schematic}
\end{figure}

We develop models capturing
the ambient temperature field evolution and spatial gradients in conjunction
with the temperature variations of the particles undergoing Brownian motion.
As a specific system, we consider particles that can become transiently trapped
within energy wells created by structures within the membrane.  We investigate
how the non-equilibrium particle diffusions impact the kinetics of escape from
the energy wells.  In our studies, the particles can diffuse into or out of
different parts of the membrane that are subject to external optical 
heating.  As the particles diffuse, they can heat up or cool down
impacting their drift-diffusion dynamics in the heterogeneous energy landscape
of the membrane.  This impacts their escape kinetics from the energy wells. 

We consider heterogeneous microstructures within the membrane 
that create energy wells of the form 
\begin{eqnarray}
\label{equ_energy_well}
\Psi(\mb{X}) = \sum_i -c_2 \exp \left[-\frac{\left(\mb{X} - \mb{X}_i
\right)^2}{2\sigma_0^2} \right].
\end{eqnarray}
The $\mb{X}_i$ form a staggered lattice as shown in
Figure~\ref{fig_hot_brownian_schematic}. 
This generates particle forces
\begin{eqnarray}
\mb{F}_X = \sum_i -\nabla_X \Psi(\mb{X}) = -c_2\left( \frac{\mb{X} -
\mb{X}_i}{\sigma_0^2}  \right) \exp \left[-\frac{\left(\mb{X} - \mb{X}_i
\right)^2}{2\sigma_0^2} \right].
\end{eqnarray}
\revB{These forces are taken to be generated by fixed microstructures of the membrane and 
the model does not involve the concentration field $q$.  The protein is treated in
the over-damped regime with mobility $\mb{M}_{XX}$ the same as in the 
previous models.}
We further consider external heating that creates a local region of 
elevated temperature within the membrane of the form
\begin{eqnarray}
\label{equ_mem_temp_hot}
\theta_m(\mb{x}) = \theta_0\left(1 + c_3 \exp \left[-\frac{\left(\mb{x} - \mb{x}_0
\right)^2}{2\sigma_3^2} \right]\right).
\end{eqnarray}
For example as induced by an external source laser source~\cite{rings2011theory}.  
We illustrate the membrane-protein system  in Figure~\ref{fig_hot_brownian_schematic}.

We perform studies of a particle initially started in the center of an energy well 
at location $\mb{X}_0$.  The initial temperature distribution in the membrane is
non-uniform given by equation~\ref{equ_mem_temp_hot}.  We consider the kinetics
of Hot Brownian Motion and how the escape time is impacted by different particle 
temperature variations $c_3$.  In particular, we study the escape time for a 
particle to diffuse to radius $r_0$ from the center $\mb{X}_0$ of the energy
well.  We perform simulations repeating this experiment for different strengths
of the energy well $c_2$ and for different amplitudes of temperature $c_3$.  The
parameters used in our simulations are shown in Table~\ref{table_energy_well}. 

\rev{We remark that the particle diffusivity is often characterized by the
Mean Square Displacement (MSD), which is based on ensemble averaging of the
particle's motions over time.  The diffusivity is then the derivative in 
time of the MSD.  In the non-equilibrium setting, this statistic is less
informative since it can exhibit more complicated non-linear behaviors 
on different time-scales as the particle moves in response to thermal gradient 
induced drifts, heats up or cools down, or 
diffuses to probe different parts of the membrane.  As an alternative, we
consider here the impact of thermal effects on the first-passage time
statistics of the non-equilibrium system.  The simulation methods
for average well escape times we report can be used for those interested 
in estimating a renormalized effective diffusivity for protein behaviors 
over larger spatial-temporal scales~\cite{ross1995stochastic,Gardiner1985}. }

\begin{table}[!h]
\setlength{\tabcolsep}{4pt} 
\global\long\def\arraystretch{1.5}
 \centering {\fontsize{7}{8}\selectfont 
\begin{tabular}{|l|l|l|l|l|l|}
\hline 
\rowcolor{atz_table1} \multicolumn{2}{|l|}{\textbf{parameter}} & \textbf{value} & \multicolumn{2}{l|}{\textbf{parameter}} & \textbf{value} \tabularnewline
\hline 
$\kappa_{PI}$  & heat conduction: particle  & \atznum{5.7e2}  & $\gamma_{p}$ & particle drag & \atznum{1e-1}\tabularnewline
\hline 
$\kappa_{CI}$ & heat conduction: interface & \atznum{3.02e3} & $C_{P}$ & specific heat: particle & \atznum{9.3e2}\tabularnewline
\hline 
$\kappa_{CC}$ & heat conduction: conc. & \atznum{2.1e-3} & $C_{C}$ & specific heat: membrane & \atznum{1.3e4}\tabularnewline
\hline 
$\kappa_{0}$ & heat conduction: fluid & \atznum{8.2e6} & $C_{I}$ & specific heat: interface & \atznum{1.4e2}\tabularnewline
\hline 
$n_{x}$ & number grid points in x & \atznum{20} & $\theta_{0}$ & baseline membrane temperature & \atznum{3.0}\tabularnewline
\hline 
$n_{y}$ & number grid points in y & \atznum{20} & $c_{2}$ & energy-well: strength & \atznum{1.5e-4}\tabularnewline
\hline 
$\Delta x$ & mesh spacing & \atznum{0.1} & $c_{3}$ & external heating strength & (varies)\tabularnewline
\hline 
$\Delta t$ & time step & \atznum{3e-3} & $X_{0}$ & energy-well: center & $[5.0/3.0,1.0]$\tabularnewline
\hline 
$k_{B}$ & Boltzmann's constant & \atznum{1e-5} & - & - & -\tabularnewline
\hline 
\end{tabular}\vspace{0.2cm}
} \caption{\rev{\textbf{Parameters for Hot Brownian Motion.} We give the values for
the SELM simulations of the particle diffusing in heterogeneous temperature
fields used in the studies for energy well escape kinetics.}}
\label{table_energy_well-1}
\end{table}

\begin{figure}[!h]
\centerline{\includegraphics[width=0.6\columnwidth]{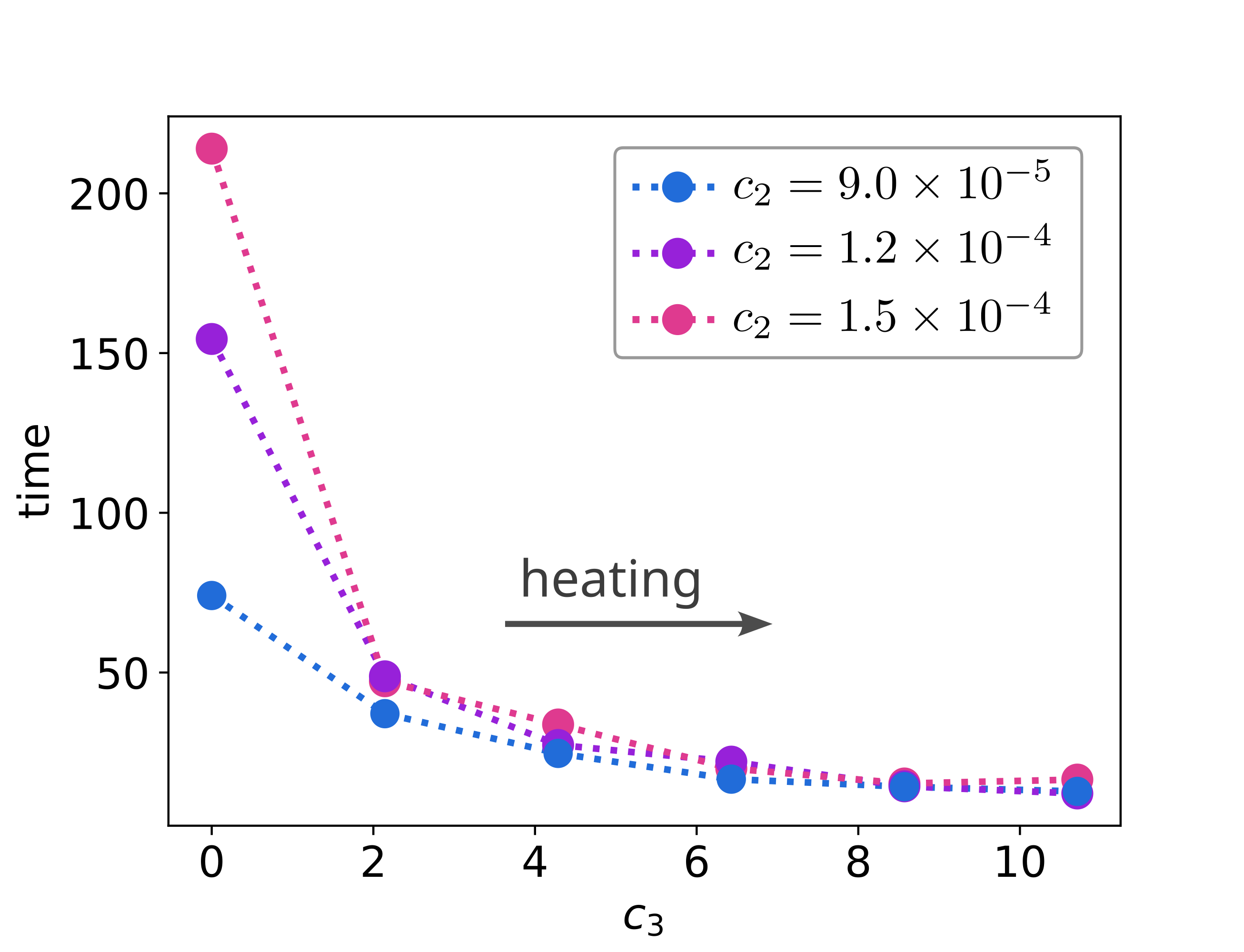}}
\caption{\rev{{\bf Hot Brownian Motion in Energy Wells.}  We show the time for
particles to escape from an energy well by diffusing to distance $r_0$ from the
well center.  The membrane has a non-uniform temperature field modulated by
$c_3$ in equation~\ref{equ_mem_temp_hot}.  This heats up the particles and can
impact the energy well escape times.  The $c_2$ gives the strength of the
energy well in equation~\ref{equ_energy_well}.  We show how the particle escape
times are impacted by the energy well strength and level of particle heating.}
}
\label{fig_hot_brownian_results}
\end{figure}

We show results in 
Figure~\ref{fig_hot_brownian_results}.  In the case of the 
strongest energy well $c_2 = \atznum{1.5e-4}$, we find 
at the baseline temperature $\theta_0 = 3.0$
with $c_3 = 0$ the particle kinetics exhibit long-duration escape times.  As the 
membrane is externally heated the particle temperature increases over time
and the diffusive motions become larger and can more readily overcome the energy
barriers.  We see as $c_3$ is increased 
these non-equilibrium diffusions have significantly smaller escape times
than the baseline constant temperature case.  When the 
external heating is large enough, the dominating time-scale becomes how long
it takes for a particle to heat up beyond a critical temperature so $k_B{T}$
is a multiple of the energy barrier size, which allows for rapid escape.  
We see the escape times become negligible as we approach 
$c_3 = 10$.  We also see only a weak dependence on the energy well 
strength $c_2$ as the external strength of heating $c_3$ increases, see 
Figure~\ref{fig_hot_brownian_results}.  

These results show some of the ways the non-equilibrium SELM approaches can be used 
to capture phenomena in the drift-diffusion dynamics of particle kinetics within  
complex heterogeneous materials that have spatially varying microstructures 
and temperature variations over time.  The results here indicate how 
trapped particles interacting with non-homogeneous temperature fields 
can impact kinetics. The SELM simulations also have the potential
to capture interesting energy exchanges and augmentations where diffusing
hot particles could locally heat up the membrane and change it locally
which could impact kinetics in future encounters with previously visited 
locations.  The temperature varying particles also provide mechanisms by which 
heat energy can be adsorbed and transferred to new locations and deposited 
through diffusion.  The kinetics involved in such mechanisms involve 
an interplay between the rate of Brownian motion, which depends on 
the particle and local temperature fields, and the rates of particle 
heating and exchanges with the environment.  The non-equilibrium SELM 
methods allow for capturing in a self-consistent manner such 
thermodynamics, kinetics, and related phenomena.

\section*{Discussion}

Biological systems involve active processes at the microstructure level that
can drive membranes and proteins into interesting regimes that are out of
thermodynamic equilibrium.  Theoretical modeling frameworks and simulation
methods were introduced for investigating non-equilibrium effects in proteins
dynamics within heterogeneous membranes.  The approaches are based on hybrid
discrete-continuum descriptions which track discrete individual proteins and
couple these to continuum fluctuating concentration and temperature fields.
This allows for investigating the roles of non-equilibrium effects in the
drift-diffusion dynamics of proteins and their coupling to spatial fields
within the membrane associated with variations in concentration and
temperature.  Since the coupling is bi-directional, this also allows for
studying exchanges of energy and other effects which impact the dynamical
evolution of both the concentration and thermal spatial fields and the
individual proteins. 

The investigations show non-equilibrium effects can play a significant role
impacting protein dynamics in mechanisms in biological systems and related
\textit{in vitro} experiments.  It was shown that both variations in
concentration of signaling molecules and their drift-diffusion kinetics can be
used to regulate spatial localization of proteins within heterogeneous membrane
structures.  It was also shown that thermal effects can play a significant role
within \textit{in vitro} experiments for probing the drift-diffusion dynamics
within the energy landscapes of heterogeneous membranes.  

The introduced approaches provide self-consistent models for studying
biophysical mechanisms involving the drift-diffusion dynamics of proteins
within heterogeneous membranes in non-equilibrium regimes.  The methods capture
the energy exchanges between the mechanical and thermal parts of the system.
It is expected these and related approaches can be used in studying diverse
types of non-equilibrium phenomena involving mechanical-thermal coupling within
biological systems and related \textit{in vitro} experiments.

\section*{Conclusion}
\rev{We have developed theory and modeling approaches for investigating the
non-equilibrium statistical mechanics of proteins immersed within heterogeneous
membranes.  We showed how these approaches could be used to obtain
self-consistent models coupling the drift-diffusion dynamics of individual
proteins  with fluctuating continuum fields for concentration and temperature
variations.  We developed numerical methods for
spatially discretizing the system and for efficiently generating the required
stochastic driving fields accounting for the fluctuations for practical
simulations.  The resulting non-equilibrium approaches were used
to investigate biological mechanisms for protein positioning and patterning
within membranes, factors in thermal gradient sensing, and kinetics of Brownian
motion of particles with temperature variations within energy landscapes of
heterogeneous membranes.  The approaches capture energy exchanges and
fluctuations between the thermal and mechanical parts of the system allowing
for investigating diverse non-equilibrium phenomena within biological systems
and materials.   This includes related applications in active soft materials,
complex fluids, and other biophysical systems.  }

\section*{Acknowledgments}
Authors were supported by supported by NSF Grant DMS-1616353 (to PJA, DJ)
and NSF Grant DMS-2306345 (to PJA, DJ). 
Author PJA also acknowledges UCSB Center for Scientific Computing NSF MRSEC
(DMR1121053) and UCSB MRL NSF CNS-1725797.  

\section*{Author Contributions}
\textit{Conceptualization:} Paul J. Atzberger;
\textit{Formal analysis:} Paul J. Atzberger, Dev Jasuja;
\textit{Investigation:} Paul J. Atzberger, Dev Jasuja;
\textit{Software:} Paul J. Atzberger;
\textit{Writing - original draft:} Paul J. Atzberger, Dev Jasuja;
\textit{Writing - review \& editing:} Paul J. Atzberger.

\newpage
\clearpage

\bibliography{cites}

\appendix
\section*{Appendix}

\section{Irreversible Operators $K^{(j)}$ and Stochastic Driving Fields
$\mb{g}^{(j)}$ for Fluctuations of the Membrane-Protein System} 
\label{appendix_K_j}
\rev{For our protein-membrane model in equation~\ref{equ_full_model}, we 
can express the irreversible processes in
the dynamics in terms of the dissipative  operators 
$\bar{K}^{(j)}$.  The protein drift-diffusion dynamics and temperature variations
corresponds to }
\begin{eqnarray}
\label{equ_K_1}
{K}^{(1)}
= 
\begin{bNiceMatrix}[last-row=3,last-col=3]
\substack{\theta_P \mb{M}_{\tiny XX}\\ \mbox{} } & -\frac{\theta_P
\mb{M}_{\tiny XX}\nabla_{\mb{X}} \mathcal{E}}{c_P} & \atzmlabel{$\mb{X}$} \\
-\frac{\nabla_{\mb{X}} \mathcal{E}^{T}\theta_P \mb{M}_{\tiny XX}}{c_P} &
\frac{\nabla_{\mb{X}}
\mathcal{E}^{T}\theta_P \mb{M}_{\tiny XX}\nabla_{\mb{X}} \mathcal{E}}{c_P^2} &
\atzmlabel{$\theta_{P}$} \\ 
\atzmlabel{$\mb{X}$} & \atzmlabel{$\theta_P$} 
\end{bNiceMatrix}.
\end{eqnarray}
We have that $\mb{F}_X = -\nabla_X \mathcal{E} = -\partial_{X} U^T$.
The concentration field diffusion and heat exchanges gives
\begin{eqnarray} 
\label{equ_K_2}
{K}^{(2)}
& = & 
\begin{bNiceMatrix}[last-row=3,last-col=3]
\atzadjustf{-\mbox{\small div}\left(\frac{q(x)\bar{\kappa}}{c_0} \nabla
\right)} & 
\mbox{\small div}
\left(
\frac{{q(x)\bar{\kappa}} \square
c_0\nabla\Phi(x) }{c_0c_C}
\right)
& \atzmlabel{$q(x)$}  \\
\frac{-c_0\nabla\Phi(x):\left({q(x)\bar{\kappa}} \nabla\right)}{c_0 c_C}  & 
\frac{c_0\nabla\Phi(x): \left({q(x)\bar{\kappa}} \square c_0\nabla\Phi(x)\right)}
{c_0 c_Cc_C} 
+ 
\frac{-\nabla\cdot \left(\tilde{\kappa_0}\theta_C^2 \nabla \right)}{c_C}
& \atzmlabel{$\theta_C$}  \\
\atzmlabel{$q(x)$} & \atzmlabel{$\theta_C$}  \\
\end{bNiceMatrix}.
\end{eqnarray}
The $\square$ denotes for the action of the operator acting on a spatial field
where to substitute the input field, such as $q(x),\theta_C(x)$. 
We have that $-\delta_q \mathcal{E} = c_0 \Phi$ from equation~\ref{equ_total_energy_grad}.
\revB{The interfacial coupling has heat exchanges that yield
\begin{eqnarray}
\label{equ_K_3}
\\
\nonumber
{K}^{(3)}
&=&
\begin{bNiceMatrix}[last-row=4,last-col=4]
\frac{\kappa_{PI}\theta_I\theta_P}{c_{P,P}}& 0 &
-\frac{\kappa_{PI}\theta_P\theta_I}{c_{P,I}} \atzmlabel{$\theta_P$} & 
\atzmlabel{$\theta_P$}  \\ 
0 & \frac{\mbox{\tiny diag}(\kappa_{CI}\dvol\;\theta_C\theta_I)}{c_{C,C} \dvol
\dvol} + \frac{K_{\mbox{\tiny heat}}}{c_{C,C} \dvol}  
& -\frac{\kappa_{CI}\dvol\;\theta_I\theta_C}{c_{C,I} \dvol} 
& \atzmlabel{$\theta_C$}  \\  
-\frac{\kappa_{PI}\theta_I\theta_P}{c_{I,P}} &
-\frac{(\kappa_{CI}\dvol\;\theta_I\theta_C)^T}{c_{I,C} \dvol} &
\frac{\kappa_{PI}\theta_P\theta_I + \theta_I \int \kappa_{CI} \theta_C
dx}{c_{I,I} } & \atzmlabel{$\theta_I$} \\
\atzmlabel{$\theta_P$} & \atzmlabel{$\theta_C$} & \atzmlabel{$\theta_I$} \\
\end{bNiceMatrix}.
\end{eqnarray}
For brevity in our notation for the operators, we show only a subset
of the rows and columns of the operators.  The other entries not shown 
are taken to be zero.}  The input and output 
degrees of freedom of the operator are 
labeled using the last row (for input entries) 
and last column (for output entries).
For example in $\bar{K}^{(1)}$, we show in
the first row the entries associated with the $\mb{X}$ degrees of
freedom and in the last row the entries associated with $\theta_P$.
We use a similar convention for the columns.
\revB{The $K_{\mbox{\tiny heat}}$ gives the heat exchange within the membrane,
given by the operator $K_{\mbox{\tiny heat}} = -\nabla \cdot \left(\kappa_C
\theta_C^2(x) \nabla\right)$.
In practice, this is approximated in our finite volume discretization approach by $\tilde{K}_{\mbox{\tiny heat}}$ with
\begin{eqnarray}
\label{equ_theta_discr}
\lbrack \tilde{K}_{\mbox{\tiny heat}}\rbrack_{(i_0,j_0),(i_0,j_0)} &=& c
\theta_{i_0,j_0}\left(\theta_{i_0+1,j_0} + \theta_{i_0-1,j_0} +
\theta_{i_0,j_0+1} + \theta_{i_0,j_0-1} \right) \\
\lbrack \tilde{K}_{\mbox{\tiny heat}}\rbrack_{(i_0\pm1,j_0),(i_0,j_0)} &=& -c\theta_{i_0,j_0}\theta_{i\pm1,j_0}, \\
\lbrack \tilde{K}_{\mbox{\tiny heat}}\rbrack_{(i,j_0\pm1),(i_0,j_0)} &=& -c\theta_{i_0,j_0}\theta_{i_0,j\pm1},
\end{eqnarray}
where $c = \kappa_C/\Delta{x}^2$ and $\theta_{i,j} = \theta_C(x_{i,j})$.  When
this operator is applied to the gradient of the entropy
$[\mathcal{D}\mathcal{S}]_{\theta_C(x)} = (c_C/\theta_C(x))\dvol$ this yields
an approximation with the same action as the Laplacian to $\theta_C$ 
which is associated with the Fourier law of heat exchange.} \rev{This provides 
for the model in equation~\ref{equ_full_model}--~\ref{equ_temperature_particle}
the key terms needed
in equation~\ref{equ_stoch_driving} to obtain the stochastic driving terms 
$\mb{g}^{(j)}$ for the fluctuations of 
the membrane-protein system. }

\section{Stochastic Field Generation Methods and \\ Factors $R^{(j)}$} 
\label{appendix_R_j}
\rev{We briefly give factorizations $R^{(j)}$ we have derived for 
generating the stochastic fields using equation~\ref{equ_h_thm}.
These can be verified to satisfy $R^{(j)}R^{(j),T} = K^{(j)}$
for $K^{(j)}$ given in equations~\ref{equ_K_1}--~\ref{equ_K_3}. 
For the particle irreversible dynamics, we have }
\begin{eqnarray}
\label{equ_B_E__particle}
R^{(1)} = \left[
\begin{array}{c}
\atzadjustf{\sqrt{\theta_P}R_M(Y)} \\
-\frac{\sqrt{\theta_P}\nabla_{X} \mathcal{E}^T R_M(Y)}{c_P}
\end{array}
\right],
\end{eqnarray}
where $R_MR_M^T = \mb{M}_{\tiny XX}(\mb{Y})$.
We generate the stochastic driving fields using
\begin{eqnarray}
\mb{h}^{(1)} = R_{1} \bsy{\xi}_{1}.
\end{eqnarray}
For the concentration field we have the factor 
\begin{eqnarray}
\label{equ_R__conc}
R^{(2)} = \left[
\begin{array}{cc}
\atzadjustf{-\mbox{\small div}\left(\sqrt{\frac{q(x)\bar{\kappa}}{c_0}}\square\right)} & 0 \\
-\frac{c_0\nabla\Phi:\sqrt{\frac{q(x)\bar{\kappa}}{c_0}}\square}{C_C} & 
-\atzadjustf{\mbox{\small div}\left(\sqrt{\frac{\bar{\kappa}_0\theta_C^2(x)}{C_C}}\square \right)}
\end{array}
\right].
\end{eqnarray}
\begin{eqnarray}
\mb{h}^{(2)} = R_{1} \bsy{\xi}_{1}.
\end{eqnarray}
For the thermal exchanges of the interface coupling, 
we break the terms down into two parts 
$K^{(3)} = K_{1}^{(3)} + \int K_{2}^{(3)}(x) dx$.
We use the factors 
\begin{eqnarray}
\label{equ_bar_K_j_M_E_part21}
\\
\nonumber
K_{1}^{(3)} &=& 
\left[
\begin{array}{ccccc}
\frac{\kappa_{PI}\theta_I\theta_P}{c_{P,P}}& 
-\frac{\kappa_{PI}\theta_P\theta_I}{c_{P,I}}\\
-\frac{\kappa_{PI}\theta_I\theta_P}{c_{I,P}} & 
\frac{\kappa_{PI}\theta_P\theta_I}{c_{I,I}}
\end{array}
\right]_{\mb{e}_P,\mb{e}_I} \\
\nonumber
&=&
\kappa_{PI}\theta_P\theta_I
\left[
\frac{1}{C_{P,P}} \mb{e}_P \mb{e}_P^T 
- 
\frac{1}{C_{P,I}} \mb{e}_P \mb{e}_I^T 
-
\frac{1}{C_{I,P}} \mb{e}_I \mb{e}_P^T 
+ 
\frac{1}{C_{I,I}} \mb{e}_I \mb{e}_I^T 
\right] \\
\nonumber
&=& R_{1}R_{1}^T,
\end{eqnarray}
This just involves the parts of the operator with 
indices corresponding to $\theta_P,\theta_I$.  We also
use denote $C_{i,j} = C_iC_j$ to keep the notation 
consistent between cases.  This has the factor 
\begin{eqnarray}
\label{equ_bar_K_j_M_E_R21}
\\
\nonumber
R_{1} &=& 
\sqrt{\kappa_{PI}\theta_I\theta_P}
\left[
\begin{array}{r}
\frac{1}{c_{P}}\mb{e}_P\\
-\frac{1}{c_{I}}\mb{e}_I
\end{array}
\right].
\end{eqnarray}
The second part has similar factorization for each spatial location since 
$\kappa_{CI} = \kappa_{CI}(x)$ with
\begin{eqnarray}
\label{equ_bar_K_j_M_E_part21}
\\
\nonumber
K_{2}^{(3)} &=& 
\left[
\begin{array}{ccccc}
\frac{\mbox{\tiny diag}\left(\kappa_{CI}\theta_I\theta_{C(x)}\;\dvol\right)}{c_{C,C} \dvol \dvol}& -\frac{\kappa_{CI}\theta_{C(x)}\theta_I\;\dvol}{c_{C,I} \dvol}\\
-\frac{\kappa_{CI}\theta_I\theta_{C(x)}\;\dvol}{c_{I,C} \dvol} & 
\frac{\int \kappa_{CI}\theta_{C(x)}\theta_I dx}{c_{I,I}}
\end{array} 
\right]_{\mb{e}_{\theta_C(x)},\mb{e}_{\theta_I}} \\
\nonumber
& = & \int \kappa_{CI}(x)\theta_{C(x)}\theta_I 
\left[\frac{1}{c_{C,C} \dvol \dvol} 
\mb{e}_{\theta_C(x)}\mb{e}_{\theta_C(x)}^T
- \frac{1}{c_{C,I} \dvol}
\mb{e}_{\theta_C(x)}\mb{e}_{\theta_I}^T\right. \\
\nonumber
&+& 
\left.
-\frac{1}{c_{I,C} \dvol} 
\mb{e}_{\theta_I}\mb{e}_{\theta_C(x)}^T
+ \frac{1}{c_{I,I}}
\mb{e}_{\theta_I}\mb{e}_{\theta_I}^T \right] dx  \\
\nonumber
&=& 
\int 
\kappa_{CI}\theta_{C(x)}\theta_I
\left[
\begin{array}{c}
\frac{1}{c_{C} \dvol} \mb{e}_{\theta_C(x)} \\
\nonumber
-\frac{1}{c_{I}} \mb{e}_{\theta_I}
\end{array} 
\right]
\left[
\begin{array}{c}
\frac{1}{c_{C} \dvol} \mb{e}_{\theta_C(x)} \\
\nonumber
-\frac{1}{c_{I}} \mb{e}_{\theta_I}
\end{array} 
\right]^T
dx \\
\nonumber
& = & \int R_{2}(x)R_{2}^T(x) dx,
\end{eqnarray}
where 
\begin{eqnarray}
\label{equ_K_j_M_E_R22}
R_{2}(x) &=& 
\sqrt{\kappa_{CI}(x)\theta_I\theta_C(x)\dvol}
\left[
\begin{array}{c}
\frac{1}{c_{C}\dvol} \mb{e}_{\theta_C(x)} \\
-\frac{1}{c_{I}} \mb{e}_{\theta_I}
\end{array} 
\right].
\end{eqnarray}
We generate the stochastic driving fields for the fluctuations using 
\begin{eqnarray}
\label{equ_M_E__R_0}
\\
\nonumber
\mb{h}^{(3)} = \mb{h}_{1} + \int \mb{h}_{2}(x)dx, \;\;
\mb{h}_{1} = R_{1} \bsy{\xi}_{1},\;\;
\mb{h}_{2}(x) = R_{2}(x) \bsy{\xi}_{2}(x).
\end{eqnarray}
Since the $\bsy{\xi}_{ij}(x_1),\bsy{\xi}_{ij}(x_2)$ have zero
correlation when $x_1 \neq x_2$, we have 
\begin{eqnarray}
\\
\nonumber
\langle 
\mb{h}_{2}
\mb{h}_{2}^T
\rangle
&=& \langle \mb{h}_{1} \mb{h}_{1}^T\rangle
+ 
\int 
\langle \mb{h}_{2} \mb{h}_{2}^T\rangle
dx \\
\nonumber
&=&  R_{1}R_{1}^T + \int  R_{2}(x)R_{2}(x)^T dx 
= K_{1}^{(3)}
+ 
\int 
K_{2}^{(3)}(x)
dx = 
K^{(3)}.
\end{eqnarray}
\rev{The expressions we have derived allow for avoiding the need to perform numerical
Cholesky factorizations each time-step.  These expressions allow for directly evaluating
operators $R^{(j)}$ to generate the stochastic driving fields needed for 
sampling the system fluctuations. }

\section{Validation of the Stochastic Numerical Methods} 
\label{appendix_validation}

\revB{We perform a few tests
to validate the stochastic numerical methods. 
This includes testing the transfer operators which are approximated 
using our spatial finite volume discretization approach in 
equations~\ref{equ_fd_grad} - \ref{equ_fd_div} and related methods discussed in
Appendix~\ref{appendix_K_j}.  We perform convergence studies 
for how our discretizations approximate the continuum operators.  
For our stochastic time-step integration methods in 
equation~\ref{eqn_stoch_num_method},
we perform tests of the generated dynamics.  This includes testing the 
spatial-temporal covariance structure for the stochastic
trajectories.  The covariance structure provides an especially useful 
test since it depends on several different parts of the stochastic 
numerical methods performing correctly.  The covariance and spatial 
approximation tests provide checks on both the 
theoretical properties of the numerical methods and
the practical implementations.}

\begin{table}[h!]
\setlength{\tabcolsep}{4pt} 
\global\long\def\arraystretch{1.5}
 \centering {\fontsize{7}{8}\selectfont 
\begin{tabular}{|l|l|l|l|l|l|}
\hline 
\rowcolor{atz_table1} \multicolumn{2}{|l|}{\textbf{parameter}} & \textbf{value}
& \multicolumn{2}{l|}{\textbf{parameter}} & \textbf{value} \tabularnewline
\hline 
$\kappa_{PI}$ & heat conduction: particle & \atznum{1.3e2} & $C_{P}$ & specific heat: particle & \atznum{1.2}\tabularnewline
\hline 
$\kappa_{CI}$ & heat conduction: interface & \atznum{1.02e2} & $C_{C}$ & specific heat: concentration & \atznum{1.4}\tabularnewline
\hline 
$\kappa_{CC}$ & heat conduction: membrane & \atznum{1.2e-2} & $C_{I}$ & specific heat: interface & \atznum{1.3e2}\tabularnewline
\hline 
$\kappa_{0}$ & heat conduction: fluid & \atznum{8.2e6} & $\theta_{0}$ & baseline membrane temperature & \atznum{3.0}\tabularnewline
\hline 
$c_{0}$ & total concentration & \atznum{1.1} & $k_{B}$ & Boltzmann's constant & \atznum{1e-5}\tabularnewline
\hline 
$n_{x}$ & number grid points in x & \atznum{5} & $\Delta x$ & mesh spacing & \atznum{0.1}\tabularnewline
\hline 
$n_{y}$ & number grid points in y & \atznum{5} & $\Delta t$ & time step & \atznum{1e-3}\tabularnewline
\hline 
\end{tabular}\vspace{0.2cm}
}
\caption{\revB{\textbf{Parameters for the Stochastic Numerical Methods.} 
We give the default values used in tests.}}
\label{table_param_validation}
\end{table}

\begin{figure}[h!]
\centerline{\includegraphics[width=0.7\columnwidth]{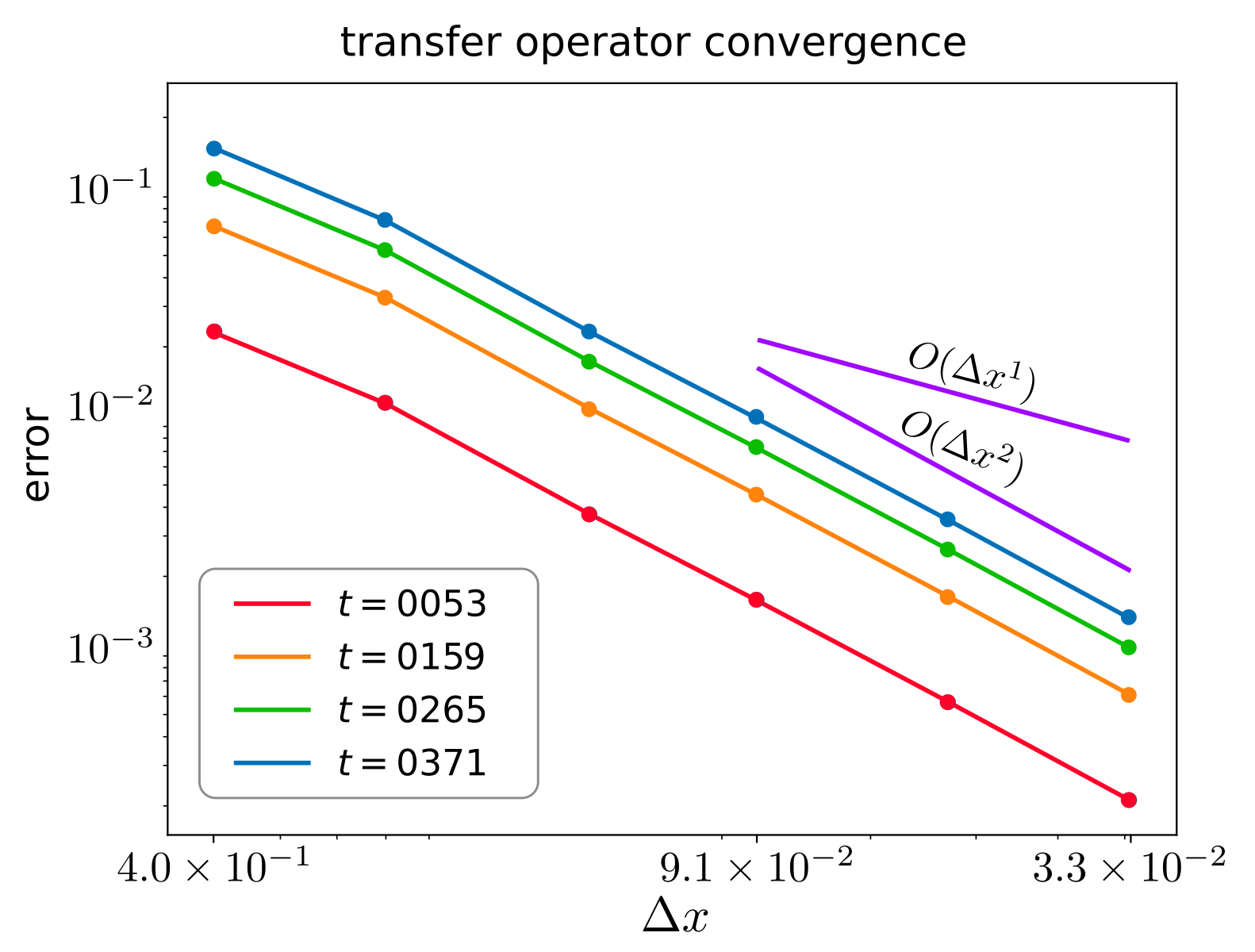}}
\caption{\revB{{\bf Transfer Operator Convergence.}  We show how the transfer operator
for the temperature field $\theta_C(x)$ converges as the spatial discretization 
$\Delta{x}$ is refined.  We test the accuracy of $\tilde{u}(\mb{x},t)$ from the numerical methods 
at different time steps $t$ using the predicted solution $u(\mb{x},t)$ 
in equation~\ref{eqn_u_predict}.  We consider the maximum error over the grid.
We find the numerical methods exhibit second-order convergence $O(\Delta{x}^2)$ 
in agreement with theory.
The spatial discretization $\Delta{x}$ becomes
smaller from left to right.  The numerical tests were performed with default
parameters in Table~\ref{table_param_validation}.  
}}
\label{fig_validation_spatial}
\end{figure}

\revB{We validate the spatial discretizations of the continuum operators
for transport within the membrane.  We investigate the accuracy 
as the spatial discretization $\Delta{x}$ for the mesh is refined.  
This impacts lateral transport of energy, such as in the temperature field 
$\theta_C$ given in equations~\ref{equ_temperature_mem} and~\ref{equ_theta_discr}.  
As a test function, we use the known analytic solution to the heat equation  
\begin{align}
\label{eqn_u_predict}
u(x_1,x_2,t) &= C_1\exp\left(-\alpha_0 t\right)\sin(2\pi k_1 x_1/L_1)\sin(2\pi k_2 x_2/L_2) + C_2 \\
\nonumber
\alpha_0 &= \frac{4\pi^2 \kappa_C}{L^2 c_C}\left(k_1^2 + k_2^2\right), &&
\end{align}
where $k_1 = k_2 = 2$, $C_1 = 3$, $C_2 = 6$.
The default parameter values used in tests are given in Table~\ref{table_param_validation}. 
We take $\kappa_{CI} = 0$ to isolate the temperature field in this test, $L_1 = L_2 = L = 2.0$,
and we vary $n_x$ with $n_y = n_x$ to refine $\Delta{x}$. 
With the initial condition at $t = 0$, we make a comparison between the
numerical solution obtained from the simulations $\tilde{u}$ with the
predicted solution $u$.  We consider the maximum error over the grid
$\epsilon = \max_{ij} |\tilde{u}(\mb{x}_{ij},t) - u(\mb{x}_{ij},t)|$ where 
$\mb{x}_{ij} = \left((i + \frac{1}{2})\Delta{x} - L/2,\;\; (j + \frac{1}{2}) \Delta{x}
- L/2\right)$, $0\leq i \leq n_x - 1$, and $0 \leq j \leq n_y$.
We show results of our convergence tests in 
Figure~\ref{fig_validation_spatial}.}

\revB{We find the numerical methods exhibit 
second-order convergence $O(\Delta{x}^2)$.  This validates the convergence 
of the numerical methods.  This also validates their accuracy and the 
scaling of their error in agreement with theory.  The results show the 
numerical methods provide accurate results both in discretizing the 
transport operators spatially and in their propagation 
over time.} 

 \begin{figure}[!h]
\centerline{\includegraphics[width=0.99\columnwidth]{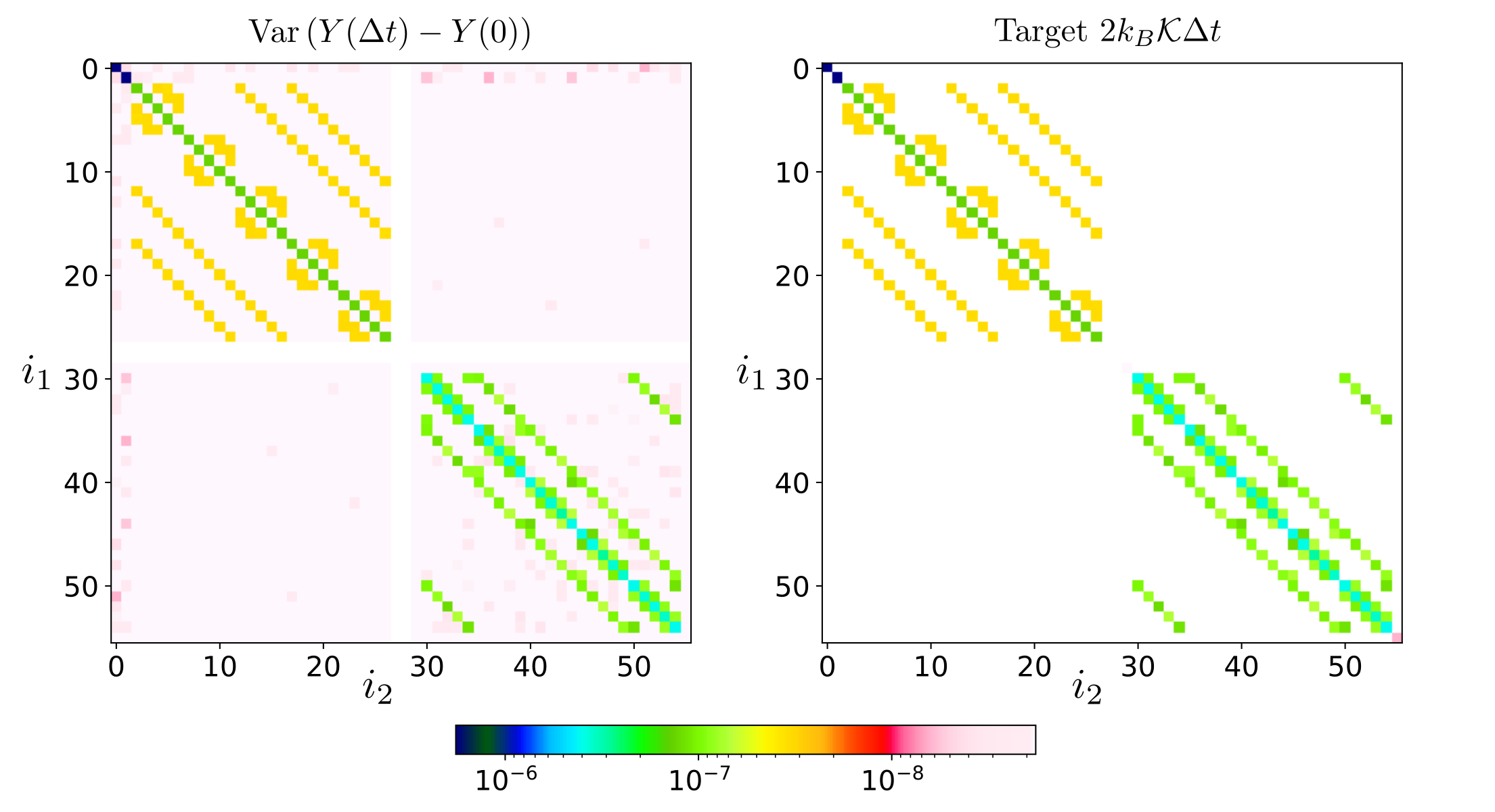}}
\caption{\revB{{\bf Covariance of Increments of the Stochastic Time-Step Integrator.}  
We show how the covariance of trajectories generated by the
stochastic numerical methods compares with the target dynamics. Results are shown 
using a log scale.  The numerical tests were performed with $n=10^4$ samples of the 
integration step with the parameters in Table~\ref{table_param_validation}.  We found 
a maximum absolute error of $\epsilon = \atznum{6.8024e-09}$.}
}
\label{fig_validation_int}
\end{figure}

\revB{We further test the stochastic numerical methods by
performing simulations of the full system dynamics $Y(t)$ for
equations~\ref{equ_full_model}-~\ref{equ_temperature_particle}. 
We use Monte-Carlo
sampling for the increments for a specified initial state $Y(0)$ to obtain
$\Delta{Y} = Y(\Delta{t}) - Y(0)$.  As discussed near equation~\ref{eqn_stoch_num_method}, our
stochastic time-step integration is a multi-stage procedure designed to capture
the drift-diffusion dynamics given in equations~\ref{equ_dY_gen}. From $n$ samples of the
stochastic trajectory generated by the numerical methods, we estimate the mean
drift contributions as $\langle \Delta{Y} \rangle = \frac{1}{n} \sum_{i=1}^n
\Delta{Y}^{(i)}$, where $\Delta{Y}^{(i)}$ is the $i^{th}$ sample.  To test the
diffusive fluctuation contributions to the dynamics, we estimate the
covariance contributions using 
\begin{eqnarray}
\mbox{var}\left( \Delta{Y}\right) = \langle
\Delta{Y}\Delta{Y}^T\rangle - \left(\langle \Delta{Y} \rangle\right)^2. 
\end{eqnarray}
We estimate the second moment using $\langle \Delta{Y}\Delta{Y}^T\rangle =
\frac{1}{n} \sum_{i=1}^n \Delta{Y}^{(i)}\Delta{Y}^{(i),T}$.  The
spatial-temporal structure of the covariance depends on several different
parts of the underlying stochastic numerical methods in order to perform correctly.  
The results also depend on our stochastic field generation methods 
based on our factorizations in Appendix~\ref{appendix_R_j}.  This 
requires that the multi-stage stochastic time integration correctly combine the
fluctuating terms while preserving the temporal contributions.  The numerical
tests were performed with $n=10^4$ samples of the integration step with the
parameters in Table~\ref{table_param_validation}.  For the initial state 
we use $\mb{X} = [5.0/3.0,1.0]$, $q(x) = 1.0$, $\theta_I = 1.2$, $\theta_P = 3.0$.
For the temperature field $\theta_C(x)$, we use the function in equation~\ref{eqn_u_predict}.
We show the results in
Figure~\ref{fig_validation_int}.  }

\revB{The empirical studies show that the stochastic numerical methods yield the correct
spatial-temporal covariance structure.  It is found that the maximum absolute error for
the tests is $\epsilon = \atznum{6.8024e-09}$.  These results show the analytic factorizations
and related implementations are working correctly to provide accurate approximations.  
The results also further show that the multi-stage stochastic integration methods 
properly handle the stochastic terms. This samples multiple sources of fluctuations 
and captures their diffusive contributions to the system dynamics.  In summary,
the results show the numerical methods and implementations are
able to provide an accurate approximation of the continuum spatial 
operators and the stochastic dynamics of the system. }

\end{document}